\documentclass[twocolumn, trackchanges]{aastex631}

\shorttitle{HESS J1813$-$178 Multi-Zone Modeling}
\shortauthors{Xian et al.}

\graphicspath{{./}{figures/}}
\usepackage{amsmath}
\usepackage{amssymb}
\allowdisplaybreaks 

\begin{document}

\title{Multi-Zone Modeling and Origin of Nonthermal Emission from HESS J1813$-$178}

\author[0009-0005-1436-6905]{Yujun Xian}
\affiliation{International Centre of Supernovae (ICESUN), Yunnan Key Laboratory of Supernova Research, School of Physics and Astronomy,Yunnan University, Kunming 650091, China; fangjun@ynu.edu.cn}

\author{Huan Yu}
\affiliation{Department of Physical Science and Technology, Kunming University, Kunming 650214, China; yuhuan.0723@163.com }

\author{Bo Wang}
\affiliation{International Center of Supernovae (ICESUN), Yunnan Key Laboratory of Supernova Research, Yunnan Observatories, Chinese Academy of Sciences, Kunming 650216, China; wangbo@ynao.ac.cn}

\author[0000-0001-8043-0745]{Jun Fang}
\affiliation{International Centre of Supernovae (ICESUN), Yunnan Key Laboratory of Supernova Research, School of Physics and Astronomy,Yunnan University, Kunming 650091, China; fangjun@ynu.edu.cn}

\begin{abstract}
HESS J1813-178 is a TeV $\gamma$-ray source with a complex multiwavelength emission structure. The source as a whole appears as an extended component, HESS J1813-178B, with an angular scale of approximately $0.7^\circ$. It also contains a compact component, HESS J1813-178A, centered on the pulsar PSR J1813-1749 and characterized by a radius of approximately $0.06^\circ$. 
To explain the broadband SED within a unified physical framework
and to reveal the physical connection between components A and B, we adopt a spatially dependent multi-zone model and describe the source as an evolving PWN-TeV halo system. Electrons are injected by the central pulsar, and their evolution under the combined effects of advection, diffusion, and energy losses is described by a Fokker-Planck-type transport equation. 
The results show that the model can reasonably reproduce the broadband SEDs of the two components.
The emission properties of component A  are consistent with a pulsar wind nebula (PWN),  whereas component B  gradually develops a halo-like morphology under suppressed diffusion. The model supports the interpretation that HESS J1813-178 is likely in a transitional stage of evolution from a PWN to a TeV halo.
\end{abstract}

\keywords{Gamma-ray sources (633) --- Pulsar wind nebulae (2215) --- High energy astrophysics (739)}

\section{Introduction}
\label{sect:1}     
PWNe are among the most representative nonthermal sources in the Galaxy. They are powered by the rotational energy loss of pulsars, which is carried outward by a relativistic wind composed mainly of high-energy particles and electromagnetic fields \citep{Martin2012, Eagle2022}.  When the pulsar wind interacts with the surrounding supernova ejecta or the interstellar medium, it produces a PWN dominated by nonthermal emission (e.g., \citealt{Kennel1984}; \citealt{Gaensler2006}; \citealt{Lu2024}; \citealt{Gong2026}).

As the system evolves, the emission morphology of a PWN does not always remain a compact nebular structure with high surface brightness. Some of the high-energy particles may gradually escape from the inner region and propagate to larger scales, thereby forming a diffuse TeV halo in the outer region \citep{Abeysekara2017science, Sudoh2019}. Compared with a traditional PWN, this component usually extends over a larger spatial scale, and its high-energy radiation is mainly produced through inverse Compton scattering of high-energy electrons off background photon fields. Recent TeV observations have shown that $\gamma$-ray halos are present around several pulsars, and that the diffusion coefficient in these regions is significantly lower than the average level in the Galaxy. This suggests that a suppressed particle transport region is likely present in the vicinity of the source (e.g., \citealt{Evoli2018}; \citealt{Albert2024}; \citealt{Adams2025}).

HESS J1813-178 is a bright Galactic TeV source with complex emission components, first detected in the Galactic $\gamma$-ray survey carried out by the High Energy Stereoscopic System (H.E.S.S.)  \citep{HESS2005Science}. 
Subsequent multiwavelength observations further revealed the counterparts and surrounding environment of this source. ASCA observations identified the X-ray counterpart AX J1813-178, VLA observations revealed the associated shell-type radio supernova remnant (SNR) G12.82-0.02, and INTEGRAL detected the hard X-ray source IGR J18135-1751 \citep{Brogan2005, Ubertini2005}. Building on these results, high-resolution X-ray observations with Chandra and XMM-Newton further resolved a compact central source and its surrounding extended nonthermal nebula, providing observational evidence for the presence of a young pulsar and its PWN in this system \citep{Funk2007, Helfand2007}. Further XMM-Newton timing observations detected 44.7 ms pulsations from PSR J1813-1749, and subsequent studies placed more precise constraints on its spin properties and detected radio pulsations from the pulsar with the Green Bank Telescope \citep{Gotthelf2009, Halpern2012, Camilo2021}. More recently, \citet{HESS2024} found that the overall emission region has an angular extent of about $0.7^\circ$ and labeled  it HESS J1813-178B, within which a compact component centered  on PSR J1813-1749 with a size of about $0.06^\circ$ was labeled  HESS J1813-178A.

Both hadronic and leptonic scenarios have been proposed to explain the multiwavelength emission from HESS J1813-178 (e.g., \citealt{Fang2010};  \citealt{Araya2018};  \citealt{Guo2024};  \citealt{Schwefer2025}).  \citet{Araya2018} suggested that the GeV  $\gamma$-ray emission may be produced by protons accelerated at the shock front of SNR G12.82-0.02. However, the hadronic scenario has difficulty accounting for the observed TeV extension, and the associated molecular cloud is also not spatially coincident with G12.82-0.02 \citep{HESS2024, Guo2024}.  On the other hand,  \citet{HESS2024}  interpreted the radiative properties of HESS J1813-178 using a one-zone leptonic model, but this scenario relies on the artificial assumption of three distinct electron populations and does not consider the spatial dependence of the particle distribution and radiative evolution.

To explain the multiwavelength emission from HESS J1813-178 in a physically self-consistent way and to reveal the relation between HESS J1813-178A and HESS J1813-178B,  we revisit the nonthermal emission properties of HESS J1813-178 using the spatially dependent multi-zone leptonic model used by \citet{Xian2026}. This model attributes the emission to the PWN powered by PSR J1813-1749 and is able to reproduce the available multiwavelength data without artificially introducing distinct electron populations. Within this multi-zone framework, we suggest that HESS J1813-178A may correspond to a PWN, whereas HESS J1813-178B may represent a TeV halo that is gradually forming in the outer region under suppressed diffusion conditions for the escaping electrons. HESS J1813-178 may be in a transitional stage from a PWN to a TeV halo.

Although the environment surrounding this source is relatively complex and contains multiple high-energy components, including a PWN and SNRs \citep{Brogan2005, Wach2023}, there is currently no clear observational evidence indicating a direct physical connection between the extended emission of HESS J1813-178 and the SNRs. To reduce the complexity of the model, we therefore adopt a multi-zone framework that follows the same physical scenario as the leptonic interpretation considered by \citet{HESS2024},  neglecting any possible radiative contribution from the SNRs.

This paper is organized as follows. Section~\ref{sect:2} briefly describes the spatially dependent multi-zone model adopted in this work. In Section~\ref{sect:3}, we present the results of fitting the model to HESS J1813-178. The summary and discussion are given in Section~\ref{sect:4}.

\section{Brief Review of the Model}
\label{sect:2}

\subsection{MODEL DESCRIPTION}
\label{sect:2.1}

In this section, we briefly introduce the spatially dependent multi-zone (multi-shell) model \citep{van2018arXiv, Xian2026}. Within this framework, high-energy electrons are injected at the termination shock, which is treated as the inner boundary, and are then transported outward through the combined effects of advection and diffusion, passing successively through the inner PWN region (Region I), and the outer diffuse TeV halo region (Region II). During transport, the particles continuously lose energy through adiabatic expansion and nonthermal radiative cooling. To simplify the calculation,  we assume that the entire PWN-TeV halo system is spherically symmetric and neglect the effect of the pulsar proper motion \citep{Lu2017ApJ, Sergio2021}. The number density distribution of high-energy particles is described by the following Fokker-Planck-type transport equation (e.g., \citealt{Vorster2013}; \citealt{Ishizaki2018}; \citealt{Kundu2024})
\begin{align}
\label{eq1}
\frac{\partial N_e}{\partial t} = & - \mathbf{V} \cdot (\nabla N_e) + \nabla \cdot \left[\kappa(r,E_e)\nabla N_e\right]\nonumber \\
& + \frac{1}{3} (\nabla \cdot \mathbf{V}) \left( \left[ \frac{\partial N_e}{\partial \ln E_e} \right] - 2 N_e \right) \nonumber \\
& + \frac{\partial}{\partial E} \left( \dot{E}_{e, \mathrm{rad}} N_e \right) + Q_{\mathrm{inj}},
\end{align}
where $N_{\rm e}(r, E_{\rm e}, t)$ represents the particle number per unit energy and per unit volume. The quantities $r$, $E_{\rm e}$, and $t$ denote the radial coordinate, particle energy, and time. $\dot{E}_{\rm e,\ rad}$ is the total radiative energy loss rate, mainly including synchrotron and inverse Compton losses.  $Q_{\rm inj}$ is the particle injection term. 
The diffusion coefficient $\kappa(r,E_e)$ is taken to be spatially uniform within each region and to retain only its energy dependence, while its normalization differs between the inner PWN region and the outer diffuse TeV halo region. Therefore, the diffusion term in Equation~(\ref{eq1}) is written in the conservative form $\nabla\cdot[\kappa(r,E_e)\nabla N_e]$.

Given that the physical environments in the PWN and TeV halo regions may differ significantly, we adopt the following piecewise form for $\kappa(r, E_{\rm e})$  \citep{Liu2019}
\begin{equation}
\label{eq2}
K(E_{\rm e}, r) =
\begin{cases}
D_1 \left( \dfrac{E_{\rm e}}{E_0} \right)^{\delta_1} & \text{if } r_0 \le r \le r_b,\\[4pt]
D_2 \left( \dfrac{E_{\rm e}}{E_0} \right)^{\delta_2} & \text{if } r_b < r \le r_{\max},
\end{cases}
\end{equation}
where $D_1$ and $D_2$ denote the normalization constants of the diffusion coefficients in the PWN and TeV halo regions, while $\delta_1$ and $\delta_2$ are the diffusion power-law indices in these two regions. To reduce the number of free parameters, we adopt  Bohm-like diffusion,  set \(\delta_1=\delta_2=\delta\), and fix \(\delta\) at 0.95  \citep{Xian2026}.  The quantities $r_0$ and $r_{\rm max}$ denote the inner and outer boundary radii of the system. The quantity $r_b$ is the boundary radius between the two regions and marks the transition from the PWN to the TeV halo.

The advection term in Equation~(\ref{eq1}) is described by the radial bulk flow velocity. We adopt the following parameterization of  \citet{van2018MN}
\begin{equation}
\label{eq3}
V(r) = V_0 \left( \frac{r}{r_0} \right)^{\alpha_V},
\end{equation}
where $V_0$ is the initial velocity at the inner boundary $r_0$, and $\alpha_V$ describes the radial dependence of the bulk flow velocity. $V_0$ is not treated as an independent fitting parameter. Following \citet{Martin2012} and \citet{Xian2026}, we adopt
\begin{equation}
\label{eq4}
V_0 = \sqrt{ \frac{10 E_0}{3 M_{\mathrm{ej}}} },
\end{equation}
where $E_0$ denotes the initial kinetic energy of the supernova ejecta, and $M_{\rm ej}$ denotes the ejecta mass. In this work, we adopt the typical values $E_0=10^{51}\,\mathrm{erg}$ and $M_{\rm ej}=8\,M_\odot$.

The pulsar continuously injects high-energy particles and energy into the system through rotational energy loss, and the corresponding injection luminosity can be written as follows  \citep{Pacini1967, Manchester1977}
\begin{equation}
\label{eq5}
L(t) = 4\pi^2 I \frac{\dot{P}}{P^3},
\end{equation}
where $P$ and $\dot{P}$ represent the pulsar's spin period and its first-order time derivative, with the pulsar's moment of inertia taken as a typical value of $I = 10^{45}\ \rm g \cdot cm^2$  \citep{Zhu2018}.  
Assuming the standard pulsar braking model $\dot{\Omega}=-K\Omega^n$, where $\Omega=2\pi/P$ is the angular velocity, $K$ is a constant, and $n$ is the braking index, the spin-down luminosity can be written as follows \citep{Pacini1973}
\begin{equation}
\label{eq6}
L(t)=L_{0}\left(1+\frac{t}{\tau_{0}}\right)^{-\frac{n+1}{n-1}}.
\end{equation}
For vacuum magnetic dipole radiation losses, the braking index is $n=3$, which is the value adopted in this work.
Here, $L_0$ denotes the pulsar's initial spin-down luminosity, and the initial spin-down timescale $\tau_0$ is given by the following equation \citep{Gaensler2006}
\begin{equation}
\label{eq7}
\tau_0 =\frac{P}{(n - 1) \dot{P}} - t_{\text{age}},
\end{equation}
where $t_{\rm age}$ is its current age.

Theoretical studies suggest that the particle injection spectrum of a pulsar can be described using various functional forms. In this work, we assume that the particle spectrum injected into the PWN-TeV halo system follows a broken power-law distribution with an exponential cutoff \citep{Holler2012, Hnatyk2022}
\begin{equation}
\label{eq8}
Q(E_{\rm e}, t) =
\begin{cases}
Q_0(t) \left( \frac{E_{\rm e}}{E_b} \right)^{-\alpha_1} & \text{if } E_{\rm min} \leq E_{\rm e} \leq E_b, \\
Q_0(t) \left[ \left( \frac{E_{\rm e}}{E_b} \right)^{-\alpha_2} \right.\times  & \\
\quad \left. \exp \left( - \frac{E_{\rm e}}{E_{\rm cut}} \right) \right] & \text{if } E_b < E_{\rm e} \leq E_{\rm max},
\end{cases}
\end{equation}
where $\alpha_1$ and $\alpha_2$ represent the particle injection spectral indices for the low-energy and high-energy regions,  with $E_b$ being the spectral break energy and $E_{\rm cut}$ the characteristic cutoff energy for the high-energy region. The normalization factor of the injection spectrum $Q_0(t)$ can be expressed by the following equation \citep{Zhang2008}
\begin{equation}
\label{eq9}
\begin{split}
\epsilon L(t) =\ & \int_{E_{\rm min}}^{E_b} Q_0(t)
\left( \frac{E_e}{E_b} \right)^{-\alpha_1} E_e \, dE_e\\ 
&+ \int_{E_b}^{E_{\rm max}} Q_0(t)
\left( \frac{E_e}{E_b} \right)^{-\alpha_2} \\
& \times \exp \left( - \frac{E_e}{E_{\rm cut}} \right) E_e \, dE_e ,
\end{split}
\end{equation}
where $E_{\rm min}$ and $E_{\rm max}$ denote the minimum and maximum particle energies.  $\epsilon$ is the efficiency with which the pulsar's spin-down luminosity is converted into particle power, and we adopt $\epsilon = 0.6$, consistent with \citet{Xian2026}.

In this model,  particle injection occurs in the innermost shell of the PWN-TeV halo system, i.e., the first shell. The particles within this shell are uniformly distributed.  Accordingly,  the injection term $Q_{\rm inj}$ in Equation~(\ref{eq1}) can be written as follows
\begin{equation}
\label{eq10}
Q_{\text{inj}} = \frac{Q(E_e, t)}{V^{1}_{\text{shell}}},
\end{equation}
where $V^{1}_{\text{shell}}$  is the volume of the first shell of the system.  After particles are injected within this shell,  they are transported outward under the combined effects of advection and diffusion,  gradually propagating to progressively outer shells.  The magnetic field distribution in the system also exhibits significant spatial dependence, and its spatiotemporal evolution can be written as follows
\begin{equation}
\label{eq11}
B(r, t) = B_{\mathrm{age}} \left( \frac{r}{r_0} \right)^{\alpha_B} \left( \frac{t}{t_{\mathrm{age}}} \right)^{\beta_B},
\end{equation}
where $B_{\rm age}$ denotes the reference magnetic field strength at $r=r_0$ and $t=t_{\rm age}$ and does not represent independently specified local magnetic fields in Region I and Region II. The magnetic field is described by the spatially and temporally dependent form $B(r,t)$, so the local magnetic field strength is jointly determined by radial position and evolutionary time.
$\alpha_B$ and $\beta_B$ are the power-law indices describing the spatial and temporal evolution of the magnetic field. 

We assume that the magnetic field is dominated by the toroidal component, while the particle bulk flow is radial. Following the ideal MHD treatment for a toroidal magnetic field and radial flow presented by \citet{Kennel1984}, and using the flux-freezing approximation adopted by \citet{van2018arXiv}, we relate the radial distributions of the magnetic field and the bulk flow speed at a given evolutionary time through
\begin{equation}
\label{eq12}
VBr = \mathrm{constant},
\end{equation}
and therefore the radial power-law indices of the magnetic field and the bulk flow velocity satisfy
\begin{equation}
\label{eq13}
\alpha_V+\alpha_B=-1.
\end{equation}
Thus, $\alpha_V$ is not treated as an independent free parameter, but is determined by the adopted value of $\alpha_B$. This relation is used to reduce the number of free parameters in the present one-dimensional phenomenological transport model rather than being derived from a fully self-consistent MHD calculation.

The negative value of $\alpha_B$ corresponds to a weak radial decrease in the magnetic field, with $B\propto r^{-0.03}$. Negative values of $\alpha_B$ have also been considered in spatially dependent PWN models \citep{van2018MN, Kundu2024}. Since the available data mainly constrain the broadband SEDs integrated over components A and B, we do not attempt to constrain a more detailed radial magnetic-field profile.

\subsection{Discretization and Numerical Solution}
\label{sect:2.2}

Using the methods adopted by \citet{van2018MN}  and \citet{Xian2026}, Equation~(\ref{eq1}) is discretized and solved numerically, from which the electron energy spectrum of the system and the corresponding SED are obtained.  The discretized grid of the model is constructed along the spatial, temporal, and particle-energy dimensions. The spatial domain is divided into multiple concentric radial shells, with the thickness of each shell defined as follows \citep{Kundu2024}
\begin{equation}
\label{eq14}
\Delta r = \frac{r_{\mathrm{max}} - r_0}{N - 1},
\end{equation}
where  \( N \) is the number of shells.  When particles reach the outer boundary of the system at \( r_{\rm max} \), they are considered to have escaped from the system.  Accordingly, the outer boundary condition is written as follows
\begin{equation}
\label{eq15}
N_e(r_{\rm max}, E_e, t) = 0,
\end{equation}
in this work, we set $r_{\max}=75.8$ pc, which approximately corresponds to the observed extent of HESS J1813-178B. This value is adopted as the numerical outer boundary in the transport calculation rather than a physical limiting radius. Since $r_{\max}$ is much larger than the central PWN region and covers the main $\gamma$-ray emission region, it helps ensure that the particle distributions in the emission regions are not artificially affected by the boundary conditions.

To prevent particles from flowing backward toward the pulsar at the termination shock,  we set the inner boundary of the system as a reflecting boundary \citep{van2018arXiv}
\begin{equation}
\label{eq16}
(N_e)_{i,j,0} = (N_e)_{i,j,1} \times \frac{\kappa/\Delta r - V_{i,j,1}/2}{\kappa/\Delta r + V_{i,j,1}/2},
\end{equation}
where  \( i \) and \( j \) are the indices of the energy and time grids,  respectively. \( (N_e)_{i,j,0} \) and \( (N_e)_{i,j,1} \) correspond to the particle numbers in the current radial spatial grid and the next adjacent radial spatial grid, respectively.  \( V_{i,j,1} \) represents the particle advection velocity in the next radial spatial grid. The energy grid is divided logarithmically and is expressed as follows \citep{Kundu2024}
\begin{equation}
\label{eq17}
\Delta E = \frac{\ln\left(\frac{E_{\mathrm{max}}}{E_{\mathrm{min}}}\right)}{M - 1},
\end{equation}
where $M$ is the number of bins in the energy grid,  and the corresponding energy boundary conditions are given by the following equation \citep{Lu2017MN}
\begin{equation}
\label{eq18}
Ne(r, E_{\rm min}, t) = 0, \quad Ne(r, E_{\rm max}, t) = 0.
\end{equation}
The time step is adaptively updated according to the Courant-Friedrichs-Lewy (CFL) stability condition \citep{CFL1928}
\begin{equation}
\label{eq19}
\Delta t = W_{\rm CFL} \cdot \min(\tau_{\rm diff}, \tau_{\rm adv}, \tau_{\rm tot})
\end{equation}
where the scaling factor $W_{\rm CFL}$ is fixed at 0.15 in this work. The quantities $\tau_{\rm diff}$, $\tau_{\rm adv}$, and $\tau_{\rm tot}$ denote the characteristic timescales for diffusion, advection, and total energy loss. Their specific forms are given by the following equations 
\begin{equation}
\label{eq20}
\tau_{\rm dif} = \frac{(\Delta r)^2}{\kappa}, \quad
\tau_{\rm adv} = \frac{\Delta r}{V}, \quad
\tau_{\rm tot} = \frac{E_e}{\left| \dot{E}_{e,{\rm tot}} \right|},
\end{equation}
where $\dot{E}_{e,{\rm tot}}$ is the total energy loss rate of the particles, including contributions from adiabatic losses and energy losses caused by nonthermal radiative processes.

After discretizing Equation~(\ref{eq1}) following the above procedure, 
we obtain the following form \citep{van2018arXiv}
\begin{align}
\label{eq21}
(1 - z + \beta)(N_e)_{i,j+1,k} &= 2 Q_{i,j,1} \Delta t \notag \\
&\;+ \frac{2}{dE_{i+1,j,k} + dE_{i,j,k}} \notag \\
&\;\quad \times \Bigg[ r_a (dE_{\mathrm{loss}})_{i+1,j,k} (N_e)_{i+1,j,k} \notag \\
&\phantom{\;\;\; \times \Bigg[}
    {} - \frac{1}{r_a} (dE_{\mathrm{loss}})_{i-1,j,k} (N_e)_{i-1,j,k} \Bigg] \notag \\
&\;+ (1 + z - \beta)(N_e)_{i,j-1,k} \notag \\
&\; + (\beta + \gamma - \eta)(N_e)_{i,j,k+1} \notag \\
&\;+ (\beta - \gamma + \eta)(N_e)_{i,j,k-1} \notag \\
&\; - 2 (\nabla \cdot \mathbf{V})_{i,j,k} \Delta t (N_e)_{i,j,k},
\end{align}
where $dE_{\rm loss}$ is the total energy loss experienced by the particles within the time step $\Delta t$. When used as a subscript, $k$ denotes the spatial grid index. The parameters $\beta$ and $\eta$ control the contribution strengths of the diffusion and advection terms,   while $\gamma$ reflects the diffusion correction in spherical geometry. $r_a$ represents the weighting correction factor introduced by the nonuniform energy grid. The parameter $z$ is a correction factor introduced after discretizing the energy loss term. 
In the numerical implementation, the diffusion coefficient is evaluated separately in the two regions, and the change in $\kappa$ at $r_b$ is treated as the interface between the two transport zones.
Within each region, $\kappa$ has no additional explicit radial dependence, and its spatial variation is represented only by the change from $D_1$ to $D_2$ at $r_b$. Therefore, this treatment should be understood as an effective two-zone approximation rather than a complete description of a diffusion coefficient that varies continuously with radius.
The definitions and detailed descriptions of the relevant physical quantities in this equation are given in \citet{Xian2026}.

\begin{deluxetable}{lll}
\tabletypesize{\footnotesize}
\tablecaption{Model parameters\label{tab1}}
\tablehead{
\colhead{Pulsar and injection parameters} & \colhead{Symbol} & \colhead{Value}
}
\startdata
Period (ms)                                         & $P$                    & 44.7 \\
Period derivative (s s$^{-1}$)                      & $\dot{P}$              & $1.26\times10^{-13}$ \\
Braking index                                       & $n$                    & 3.0 \\
Distance (kpc)                                      & $d$                    & 6.2 \\
Characteristic age (kyr)                            & $\tau_{\rm c}$         & 5.6 \\
Present-day spin-down luminosity (erg s$^{-1}$)     & $L_{\mathrm{age}}$     & $5.6\times10^{37}$ \\
Particle energy conversion                          & $\epsilon$             & 0.6 \\
\sidehead{Diffusion parameters}
Diffusion normalization in Region I                 & $D_{1}$                & $1.0 \times 10^{25}$ \\
Diffusion normalization in Region II                & $D_{2}$                & $1.0 \times 10^{26}$ \\
Diffusion index                                     & $\delta$               & 0.95 \\
\sidehead{Target photon parameters}
CMB temperature (K)                                 & $T_{\mathrm{CMB}}$     & 2.72 \\
CMB energy density (eV cm$^{-3}$)                   & $U_{\mathrm{CMB}}$     & 0.26 \\
IR temperature (K)                                  & $T_{\mathrm{IR}}$      & 35.0 \\
IR energy density (eV cm$^{-3}$)                    & $U_{\mathrm{IR}}$      & 1.0 \\
Starlight temperature (K)                           & $T_{\mathrm{st}}$      & 3000.0 \\
Starlight energy density (eV cm$^{-3}$)             & $U_{\mathrm{st}}$      & 1.5 \\
\sidehead{Fitted parameters}
Low-energy particle injection index                 & $\alpha_1$             & 1.8 \\
High-energy particle injection index                & $\alpha_2$             & 2.9 \\
Break energy (erg)                                  & $E_{\rm b}$            & 2.5 \\
Cutoff energy (erg)                                 & $E_{\rm cut}$          & $1.0\times 10^{2}$ \\
Present-day magnetic field ($\mu$G)                 & $B_{\rm age}$          & 6.5 \\
\enddata
\end{deluxetable}

\begin{figure}[t]
\centering
\gridline{\fig{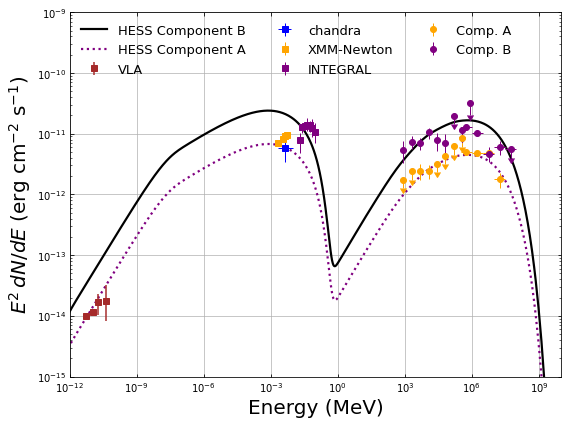}{0.45\textwidth}{}}
\caption{Comparison of the broadband SEDs of HESS J1813-178 with observational data. 
The purple dashed line represents the broadband SED of component A ($r\leq0.06^\circ$), corresponding to Region I in the model. The black solid line represents the broadband SED of component A ($r\leq0.7^\circ$), including the cumulative emission contributions from Region I and Region II.
The radio data are from VLA measurements \citep{Brogan2005},  the X-ray data are from Chandra \citep{Helfand2007}, XMM-Newton \citep{Funk2007}, and INTEGRAL \citep{Ubertini2005}, and the high-energy  $\gamma$-ray data are from H.E.S.S. observations \citep{HESS2024}.}
\label{fig1}
\end{figure}   

\begin{figure}
\centering
\gridline{\fig{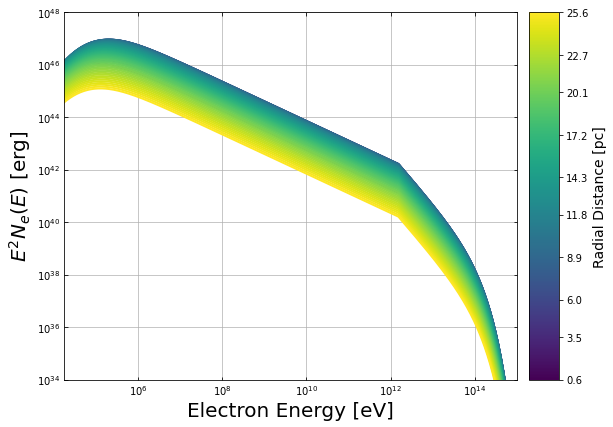}{0.45\textwidth}{}}
\gridline{\fig{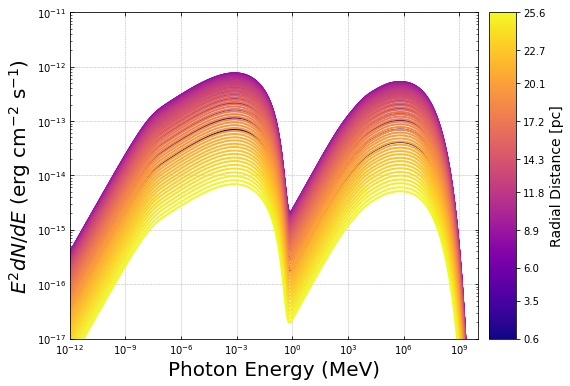}{0.45\textwidth}{}}
\caption{Upper panel: the electron spectra at different radial distances from the center outward. Lower panel: the corresponding SEDs at the same radial distances.}
\label{fig2}
\end{figure}

\begin{figure*}  
\centering
\gridline{\fig{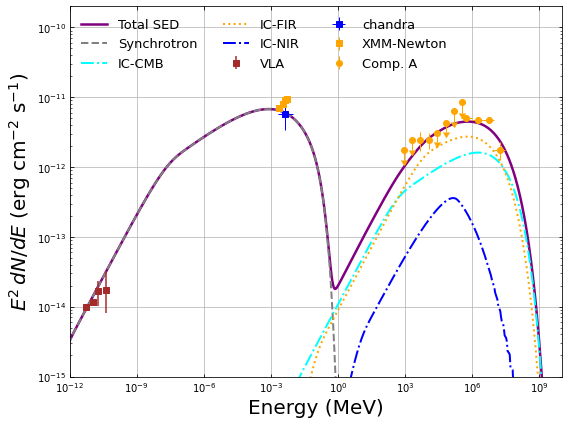}{0.45\textwidth}{}
          \fig{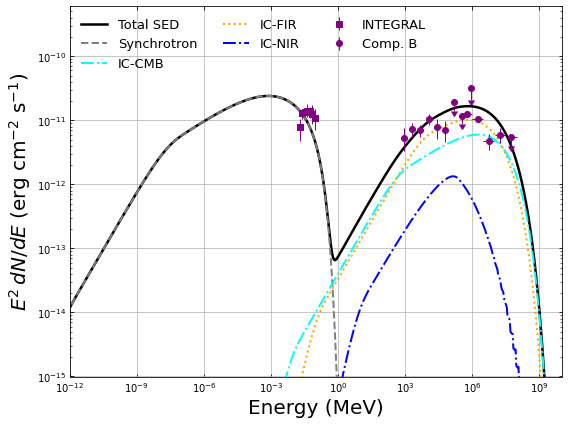}{0.45\textwidth}{}}
\caption{Left panel: the different nonthermal emission components of HESS J1813-178 A ($r \leq 0.06^\circ$).  Right panel: the different nonthermal emission components of HESS J1813$-$178 B ($r \leq 0.7^\circ$). These components include synchrotron radiation (gray dashed line), inverse Compton emission from the scattering of CMB photons (cyan dash-dotted line), inverse Compton emission from the IR photon field (orange dotted line), and inverse Compton emission from the st photon field (blue dash-dotted line).}
\label{fig3}
\end{figure*}

\begin{figure}
\centering
\gridline{\fig{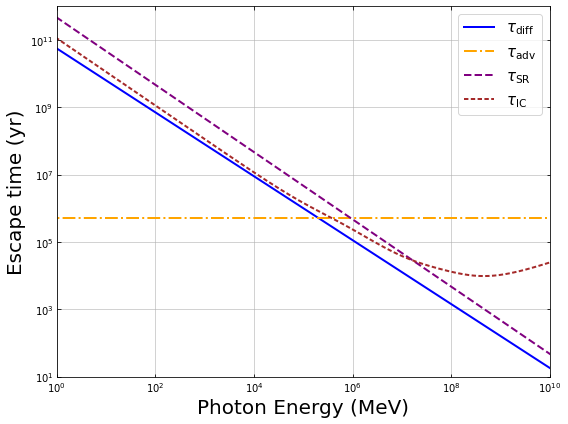}{0.45\textwidth}{}}
\caption{Cooling times for different loss mechanisms at $t = 5.6\,\mathrm{kyr}$ and $r = 0.7^\circ$ (diffusion, advection, synchrotron and inverse Compton).}
\label{fig4}
\end{figure}

\begin{figure*}  
\centering
\gridline{\fig{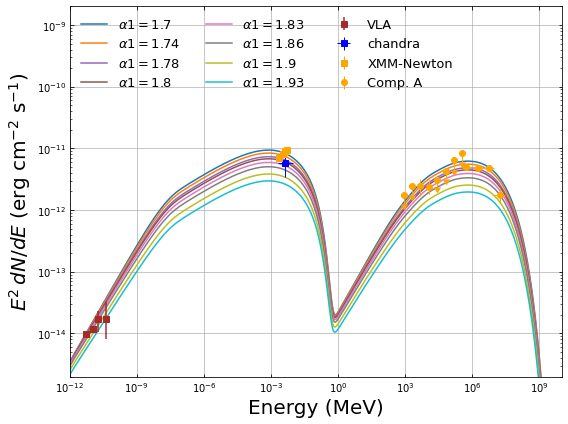}{0.45\textwidth}{}
          \fig{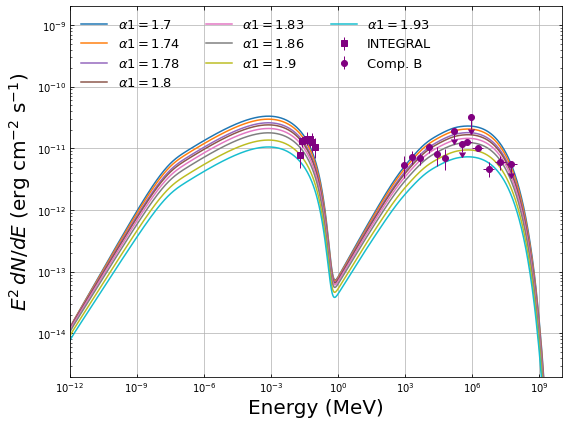}{0.45\textwidth}{}}
\caption{Effect of the low-energy particle injection index $\alpha_1$ on the region-resolved SEDs. Left panel: component A for $r \leq 0.06^{\circ}$.  Right panel: component B for $r \leq 0.7^{\circ}$.}
\label{fig5}
\end{figure*}

\begin{figure*}  
\centering
\gridline{\fig{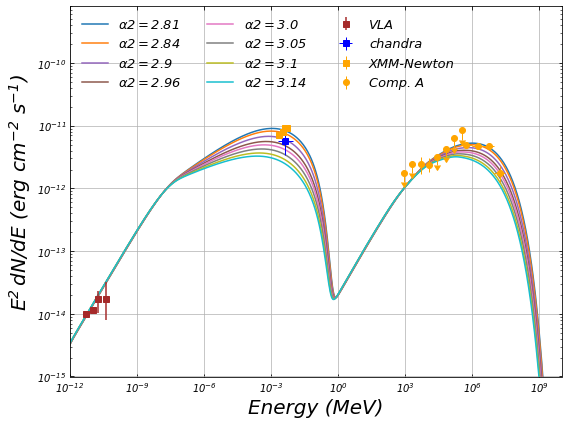}{0.45\textwidth}{}
          \fig{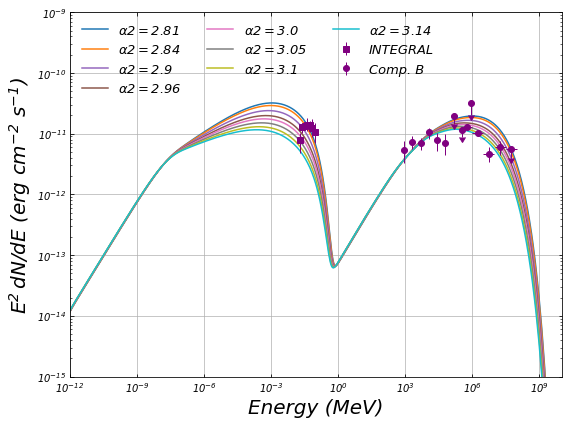}{0.45\textwidth}{}}
\caption{Effect of the high-energy particle injection index $\alpha_2$  on the region-resolved SEDs. Left panel: component A for $r \leq 0.06^{\circ}$.  Right panel: component B for $r \leq 0.7^{\circ}$.}
\label{fig6}
\end{figure*}

\begin{figure*}  
\centering
\gridline{\fig{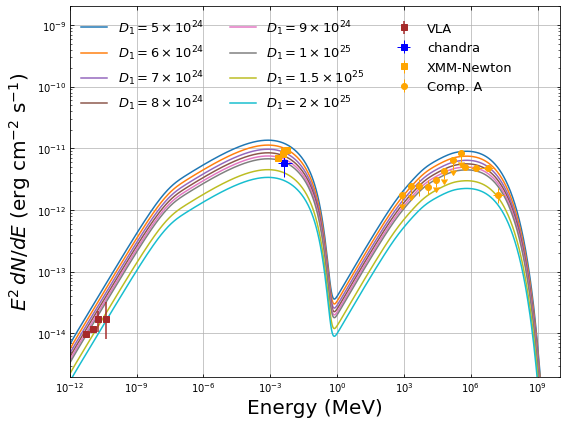}{0.45\textwidth}{}
          \fig{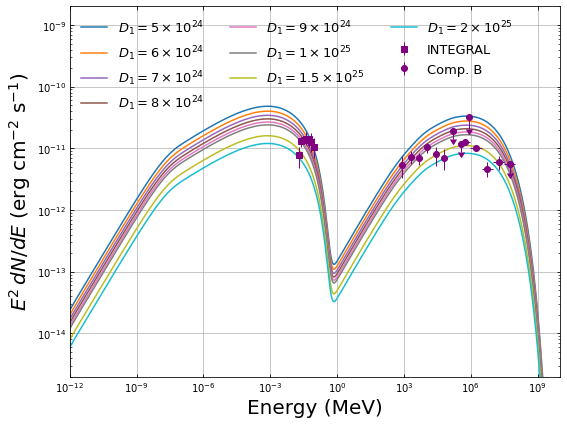}{0.45\textwidth}{}}
\caption{Effect of the diffusion normalization coefficient $D_1$ on the region-resolved SEDs. Left panel: component A for $r \leq 0.06^{\circ}$.  Right panel: component B for $r \leq 0.7^{\circ}$.}
\label{fig7}
\end{figure*}

\begin{figure}
\centering
\gridline{\fig{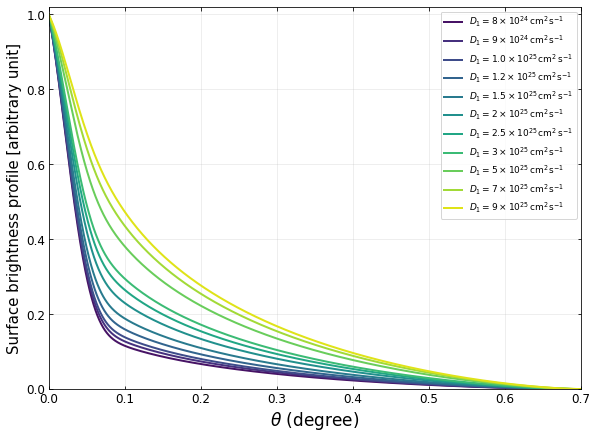}{0.45\textwidth}{}}
\caption{Effect of the diffusion normalization coefficient $D_1$ on the normalized projected radial surface brightness distribution}
\label{fig8}
\end{figure}

\begin{figure}
\centering
\gridline{\fig{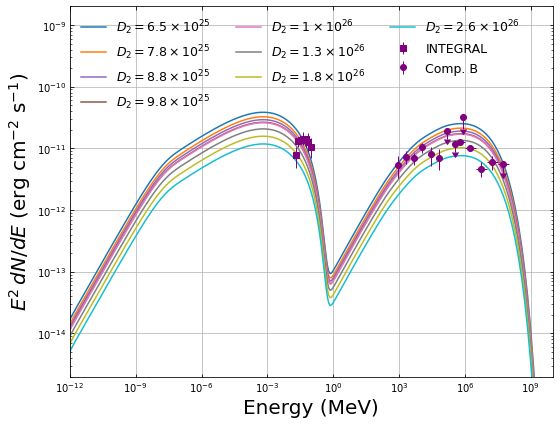}{0.45\textwidth}{}}
\caption{Effect of the diffusion normalization coefficient $D_2$ on the broadband SEDs of component B ($r \leq 0.7^\circ$)}
\label{fig9}
\end{figure}

\begin{figure*}  
\centering
\gridline{\fig{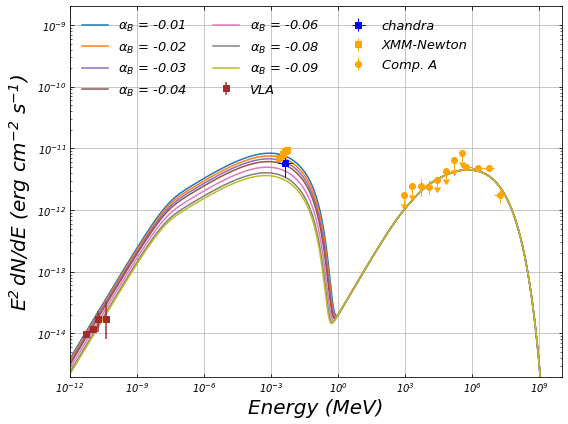}{0.45\textwidth}{}
          \fig{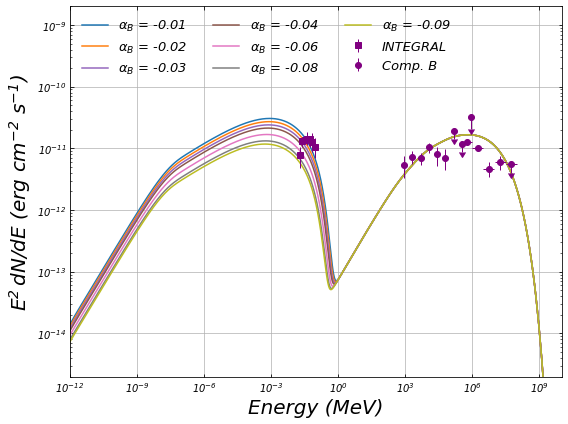}{0.45\textwidth}{}}
\caption{Effect of the radial magnetic-field index $\alpha_B$ on the region-resolved SEDs. Left panel: component A for $r \leq 0.06^{\circ}$.  Right panel: component B for $r \leq 0.7^{\circ}$.}
\label{fig10}
\end{figure*}

\section{Model Fitting}
\label{sect:3}

In this section, we apply the model described above to the broadband SED of HESS J1813-178. The characteristic angular scales measured by H.E.S.S. are used to define the corresponding emission integration regions, while the radio shell, detailed X-ray morphology, and two-dimensional $\gamma$-ray morphology are not directly modeled.
The multiwavelength nonthermal emission from this source can be broadly divided into two energy bands: the low-energy band is dominated by synchrotron radiation, whereas the GeV--TeV emission is mainly produced by inverse Compton scattering of high-energy electrons \citep{Wu2023, Tian2023}.  According to   \citet{Fang2010}, we consider the cosmic microwave background (CMB), infrared (IR), and starlight (st) fields as the target photon fields for inverse Compton scattering in HESS J1813-178. Their temperatures and energy densities are set to $T_{\rm CMB}=2.72~{\rm K}$ and $U_{\rm CMB}=0.26~{\rm eV~cm^{-3}}$, $T_{\rm IR}=35.0~{\rm K}$ and $U_{\rm IR}=1.0~{\rm eV~cm^{-3}}$, and $T_{\rm st}=3000.0~{\rm K}$ and $U_{\rm st}=1.5~{\rm eV~cm^{-3}}$.
These photon field parameters are adopted as fixed inputs rather than free parameters in the model fitting. The same target photon field parameters are applied to both Region I and Region II.

Observations indicate that the central pulsar of HESS J1813$-$178 is PSR J1813-1749, which is among the Galactic pulsars with the highest spin-down power \citep{Halpern2012}. This pulsar was first discovered through X-ray observations with Chandra  \citep{Camilo2021}. It has a spin period of $P=44.7~{\rm ms}$,  a period derivative of $\dot{P}=1.26\times10^{-13}~{\rm s~s^{-1}}$ \citep{Camilo2021}.  The present-day spin-down luminosity is adopted as $L_{\rm age}=5.6\times10^{37}\ {\rm erg\,s^{-1}}$, following the timing measurements reported by \citet{Gotthelf2009} and  \citet{Halpern2012}.
The true age of the system was estimated to be approximately \(4.7\) kyr by \citet{HESS2024}. Therefore, in this work, we adopt \(t_{\rm age}=4.7\) kyr. 

The distance to this source remains uncertain.  \citet{Camilo2021} showed that, depending on the Galactic free-electron density model adopted, the measured dispersion measure may correspond to a distance of $d=6.2$ kpc or $d=12$ kpc. The H.E.S.S. spectral data and spatial definitions used in this work to constrain components A and B are taken from \citet{HESS2024}. That study performed a unified spectral and morphological analysis of the two components and subsequently adopted $d=6.2$ kpc in its leptonic model for HESS J1813-178. To remain consistent with the observational inputs adopted in this work, we use $d=6.2$ kpc as the baseline distance.

Nevertheless, the larger distance of approximately 12 kpc cannot currently be excluded. If a larger distance is adopted, the same observed angular scales would correspond to larger physical scales, while a higher intrinsic luminosity would be required to reproduce the same observed flux. Therefore, model parameters related to the particle injection energy budget and spatial transport, such as the particle injection normalization, diffusion coefficients, and other parameters associated with spatial scales, would need to be adjusted accordingly.

Using the parameters listed in Table~\ref{tab1}, we assume that the particles injected into the system follow a broken power-law distribution with an exponential cutoff as given by Equation~(\ref{eq8}). We then use the particle spectrum evolved to $t=t_{\rm age}$ as input to the Python package Naima to calculate the SED \citep{Zabalza2015}, which is shown in Figure~\ref{fig1}. The results show that this model can simultaneously account for the emission properties of components A and B of HESS J1813-178 within a unified framework of electron injection and transport.

As shown in Figure~\ref{fig1}, the model curve for HESS J1813-178A can reasonably reproduce its radio and X-ray emission, as well as the $\gamma$-ray  emission of the compact component A observed by \citet{HESS2024}, and exhibits the typical multiwavelength emission characteristics of a PWN \citep{Torres2014, Wu2025}. Therefore, this part of the emission can be naturally interpreted as originating from the PWN powered by PSR J1813-1749. In contrast, the broadband SED of HESS J1813-178B is more extended and shows stronger emission in the high-energy $\gamma$-ray band. This may be because escaping electrons undergo suppressed diffusion in the diffuse region surrounding the PWN, allowing them to remain there for a longer time and continue to produce inverse Compton emission by scattering background photons. This suggests that HESS J1813-178 B is likely in a stage of evolution toward a TeV halo. The overall fitting results indicate that the multiwavelength emission of HESS J1813-178 can be consistently explained by a continuously evolving PWN-TeV halo system, without invoking multiple independent ad hoc electron populations.

To illustrate the spatial dependence of the multi-zone model more clearly,  Figure~\ref{fig2}  shows the electron spectra in different radial shells and their corresponding SEDs. 
For clarity, only the radial shells within $r\leq25.6$ pc are shown, since plotting all shells would make the figure overly crowded and reduce its readability. The full particle transport calculation extends to the numerical outer boundary $r_{\max}=75.8$ pc.
Both the electron spectra and their radiative SEDs exhibit clear radial evolution. As the particles are transported outward from the inner region of the system,  their overall energy distribution gradually decreases, with a more pronounced decline at the high-energy end.

In the present multizone model, the electron spectrum in each radial shell results from the cumulative effects of continuous injection, advection, energy-dependent diffusion, and radiative losses. Therefore, electrons in the same shell have different injection times and propagation histories, so the radial distance cannot be associated with a unique cooling age. In addition, the energy-dependent diffusion coefficient and the spatially and temporally dependent magnetic field $B(r,t)$ lead to different transport histories and cumulative cooling histories. Consequently, the cooling effect is mainly manifested as a gradual weakening and softening of the high-energy end of the electron spectrum with increasing radius, rather than as the formation of a clear and sharp cooling break.

Figure~\ref{fig3} shows the contributions of different physical processes to the emission from components A and B of HESS J1813-178. The left panel corresponds to HESS J1813-178A, integrated within $r\leq0.06^\circ$, whereas the right panel corresponds to HESS J1813-178B, integrated within $r\leq0.7^\circ$. The differences between the two panels mainly arise from the different spatial integration ranges and the corresponding spatial distributions of electrons.
At low energies, the emission is dominated by synchrotron radiation, whereas inverse Compton scattering dominates in the TeV $\gamma$-ray band. Among the background photon fields, the cosmic microwave background and infrared fields dominate the high-energy emission, while the starlight field makes only a minor contribution.

Figure~\ref{fig4}  presents the characteristic timescales associated with different cooling and transport processes at $t=t_{\rm age}$. For low-energy particles, the diffusion and radiative-cooling timescales are longer than the advection timescale, indicating that their transport is mainly governed by advection. 
These low-energy electrons can therefore be transported to larger radial distances before undergoing significant radiative cooling, so the low-energy part of the electron spectrum varies relatively slowly with radius. With increasing particle energy, both the diffusion and radiative cooling timescales decrease significantly, and radiative cooling, especially synchrotron cooling, becomes increasingly important at high energies. As a result, high-energy electrons lose energy more rapidly during outward propagation, causing the high-energy end of the electron spectrum to gradually weaken and soften with increasing radius.

In terms of spatial morphology, the synchrotron emission produced by low-energy electrons is expected to have a relatively large spatial extent. In contrast, the propagation distance of high-energy electrons is limited by stronger radiative cooling, so high-energy synchrotron emission, especially X-ray emission, is expected to be more concentrated in the inner region and to have a smaller spatial scale than radio emission. For TeV $\gamma$-ray emission, the spatial distribution is also affected by both energy-dependent diffusion and radiative cooling.

In this work, the diffusion normalization adopted for Region II is $D_2=1.0\times10^{26}~{\rm cm^2~s^{-1}}$, which is about two orders of magnitude lower than the canonical diffusion coefficient typically used for particle transport in the Galactic interstellar medium \citep{Ptuskin2006, Strong2007}. This result indicates that high-energy particles in this region are subject to significant diffusion suppression, allowing them to remain in the diffuse region surrounding the PWN for a longer time. This provides favorable conditions for the extended emission region to evolve toward a TeV halo.

To investigate the effects of the injection spectral indices on the SED, we explore the parameter space of $\alpha_1$ and $\alpha_2$.  Figure~\ref{fig5} shows the SED variations of HESS J1813-178A and HESS J1813-178B for different values of $\alpha_1$. The results show that, as $\alpha_1$ increases,  the low-energy electron spectrum becomes softer, leading to an overall decrease in the synchrotron emission from the radio to X-ray bands, together with a certain reduction in the inverse Compton emission in the GeV-TeV band. This indicates that the low-energy electron population not only determines the synchrotron intensity, but also affects the high-energy $\gamma$-ray output by modifying the overall electron energy distribution.  Figure~\ref{fig6} presents the effects of different values of $\alpha_2$ on the SEDs of components A and B.   Compared with $\alpha_1$, $\alpha_2$ has a weaker influence on the radio emission at the low-energy end of the SED and mainly affects the peak amplitudes of the synchrotron and inverse Compton components. Based on the comparison of these parameter sets,  we find that the model with $\alpha_1=1.8$ and $\alpha_2=2.9$ can simultaneously reproduce the multiwavelength observations of HESS J1813-178 A and HESS J1813-178 B reasonably well,  supporting a unified interpretation of this source within a spatially dependent multi-zone model.

To further examine the spatial and spectral properties predicted by the model, we first calculated the SEDs of components A and B for different values of $D_1$, as shown in Figure~\ref{fig7}. Furthermore, as shown in Figure~\ref{fig8}, following the method proposed by \citet{van2018MN}, we integrated the emissivity of each radial shell along the line of sight and calculated the projected radial surface-brightness profiles in the 6-50 TeV energy range for different values of $D_1$.  \citet{HESS2024} reported the characteristic spatial extent of HESS J1813-178 but did not provide a radial surface-brightness profile suitable for direct point-by-point comparison. Therefore, we use normalized radial surface-brightness profiles to examine the relative effects of variations in the transport parameters on the spatial distribution rather than to perform a quantitative fit to the complete two-dimensional $\gamma$-ray morphology. The results show that variations in $D_1$ affect not only the broadband emission spectrum but also the radial extent of the TeV emission. For $D_2$, which primarily controls particle diffusion in Region II, we further examined its effect on the SED of component B, as shown in Figure~\ref{fig9}.

In addition, as shown in Figure~\ref{fig10}, we investigated the effect of the radial index of the magnetic field, $\alpha_B$, on the SEDs of components A and B. In the present model, the radial indices of the magnetic field and bulk flow velocity satisfy the relation given by Equation~(\ref{eq13}), and $V_0$ is not an independent fitting parameter. Therefore, we did not perform a separate scan of the advection velocity. Under this constraint, varying $\alpha_B$ also changes the radial profile of the bulk flow velocity, allowing us to further examine how changes in the radial profiles of the magnetic field and bulk flow velocity affect the model results.

\section{Summary and Discussion}
\label{sect:4}

Based on the multi-zone model adopted by \citet{Xian2026},   we present a unified modeling and interpretation of the multiwavelength nonthermal emission from HESS J1813-178.
The model adopts a single electron population continuously injected by the central pulsar and follows its spatial evolution under the combined effects of advection, energy-dependent diffusion, and radiative cooling.

Component A corresponds to the compact integration region within $r\leq0.06^\circ$, while component B corresponds to the larger integration region within $r\leq0.7^\circ$, which contains component A (\citealt{HESS2024}). Since the same particle injection framework and target photon fields are adopted, the different radiation properties of the two components naturally arise from the different spatial distributions and spectral evolution of electrons within the corresponding integration regions. Therefore, the PWN-like emission of component A and the more extended TeV emission of component B can be explained within the same spatially dependent multi-zone transport framework, without invoking multiple independent particle populations.

Some model parameters are fixed according to observational constraints rather than treated as free parameters in the fitting. For example, the source distance, pulsar age, and external photon field parameters are adopted from previous studies, while the particle injection, magnetic field, and transport parameters are constrained by the broadband emission of HESS J1813-178.

We note that some degeneracies exist among the model parameters. For example, the magnetic field strength and electron normalization affect the synchrotron and inverse Compton emission, while the diffusion coefficient and the advection velocity influence the electron spatial distribution. Therefore, the derived parameters represent a physically consistent solution rather than a unique determination. The adopted parameters are also within the typical ranges used in spatially dependent models of young PWNe and PWN$-$TeV halo systems (e.g., \citealt{van2018MN}; \citealt{Xian2026}).

The fitting results show that, for reasonable parameter values, the model can reasonably reproduce both the compact  component A and the extended component B  of HESS J1813-178. The emission properties of  component A exhibit typical PWN characteristics. In contrast, component B  shows a more extended structure with enhanced TeV emission. This suggests that component B  may originate from escaped particles in the diffuse region surrounding the PWN, where suppressed diffusion may facilitate the gradual formation of a TeV halo. We therefore suggest that HESS J1813-178 likely represents a transitional stage in the evolution from a PWN to a TeV halo.

The one-dimensional, spherically symmetric multizone transport model adopted in this work describes the spatial evolution of electrons under the effects of advection, diffusion, and radiative losses, and explains the broadband emission properties of different spatial regions. However, this model still includes some necessary simplifications. For example, we assume a spherically symmetric structure and neglect the possible contribution from the surrounding SNR environment. In addition, this work mainly focuses on the spatially dependent particle transport processes and the broadband SEDs integrated over different spatial regions, rather than on  detailed multidimensional morphological modeling.

Future model improvements should not be evaluated solely based on broadband SED fitting. A more comprehensive model should simultaneously account for spatially resolved observations, including energy-dependent $\gamma$-ray morphology, X-ray surface brightness distributions, broadband spectral energy distributions, and possible contributions from the surrounding SNR environment within a unified transport framework.

Another uncertainty in the present model arises from the source distance. Since the distance is not treated as a free parameter or independently constrained in this work, the model parameters reported here should be understood as representative values for the adopted fiducial distance of $d=6.2$ kpc. A larger distance would imply larger physical scales and a higher intrinsic luminosity, which would affect the quantitative constraints on the particle energy budget and transport parameters. More precise distance measurements in the future will help further constrain the physical size, energy requirements, and associated particle transport parameters of this source.

\section*{acknowledgements}
We sincerely thank the anonymous referee for their constructive comments, which helped us improve the manuscript.
JF is supported by the National Natural Science Foundation of China (grant Nos. 12563010 and 12393852), the Yunnan Fundamental Research Projects (grant No. 202501AS070068), and the International Centre of Supernovae (ICESUN) of the Yunnan Key Laboratory of Supernova Research (grant No. 202505AV340004). BW is supported by the National Natural Science Foundation of China (No 12225304), the Yunnan Revitalization Talent Support Program (Yunling Scholar Project), the Yunnan Science and Technology Program (Nos  202605AS350010 and 202601BC070011). Y.J.Xian is supported by the Graduate Research Innovation Fund of Yunnan University (grant No. KC-262615769).

\bibliographystyle{aasjournal}
\bibliography{sample631}   

@ARTICLE{Martin2012,
       author = {{Mart{\'\i}n}, Jonatan and {Torres}, Diego F. and {Rea}, Nanda},
        title = "{Time-dependent modelling of pulsar wind nebulae: study on the impact of the diffusion-loss approximations}",
      journal = {\mnras},
         year = 2012,
        month = nov,
       volume = {427},
       number = {1},
        pages = {415-427},
          doi = {10.1111/j.1365-2966.2012.22014.x},
archivePrefix = {arXiv},
       eprint = {1209.0300},
 primaryClass = {astro-ph.HE},
       adsurl = {https://ui.adsabs.harvard.edu/abs/2012MNRAS.427..415M}
}

@ARTICLE{Eagle2022,
       author = {{Eagle}, Jordan and {Castro}, Daniel and {Temim}, Tea and {Ballet}, Jean and {Slane}, Patrick and {Gelfand}, Joseph and {Kerr}, Matthew and {Ajello}, Marco},
        title = "{MeV-GeV Gamma-Ray Emission from SNR G327.1-1.1 Discovered by the Fermi-LAT}",
      journal = {\apj},
         year = 2022,
        month = dec,
       volume = {940},
       number = {2},
          eid = {143},
        pages = {143},
          doi = {10.3847/1538-4357/ac9eb4},
archivePrefix = {arXiv},
       eprint = {2211.10281},
 primaryClass = {astro-ph.HE},
       adsurl = {https://ui.adsabs.harvard.edu/abs/2022ApJ...940..143E}
}

@ARTICLE{Kennel1984,
       author = {{Kennel}, C.~F. and {Coroniti}, F.~V.},
        title = "{Confinement of the Crab pulsar's wind by its supernova remnant.}",
      journal = {\apj},
         year = 1984,
        month = aug,
       volume = {283},
        pages = {694-709},
          doi = {10.1086/162356},
       adsurl = {https://ui.adsabs.harvard.edu/abs/1984ApJ...283..694K}
}

@ARTICLE{Gaensler2006,
       author = {{Gaensler}, Bryan M. and {Slane}, Patrick O.},
        title = "{The Evolution and Structure of Pulsar Wind Nebulae}",
      journal = {\araa},
         year = 2006,
        month = sep,
       volume = {44},
       number = {1},
        pages = {17-47},
          doi = {10.1146/annurev.astro.44.051905.092528},
archivePrefix = {arXiv},
       eprint = {astro-ph/0601081},
 primaryClass = {astro-ph},
       adsurl = {https://ui.adsabs.harvard.edu/abs/2006ARA&A..44...17G}
}

@ARTICLE{Lu2024,
       author = {{Lu}, Fang-Wu and {Zhu}, Bo-Tao and {Hu}, Wen and {Zhang}, Li},
        title = "{Turbulent Transport Characteristics of the Particles within Pulsar Wind Nebulae 3C58 and G54.1+0.3}",
      journal = {\apj},
         year = 2024,
        month = dec,
       volume = {977},
       number = {2},
          eid = {240},
        pages = {240},
          doi = {10.3847/1538-4357/ad946f},
       adsurl = {https://ui.adsabs.harvard.edu/abs/2024ApJ...977..240L}
}

@ARTICLE{Gong2026,
       author = {{Gong}, Yunlu and {Zhou}, Liancheng and {Xia}, Qi and {Zhang}, Li and {Fang}, Jun},
        title = "{Unravelling the nature of the TeV source HESS J1857+026 with a pulsar wind nebula scenario}",
      journal = {\mnras},
         year = 2026,
        month = apr,
       volume = {547},
       number = {4},
          eid = {stag403},
        pages = {stag403},
          doi = {10.1093/mnras/stag403},
       adsurl = {https://ui.adsabs.harvard.edu/abs/2026MNRAS.547ag403G}
}

@ARTICLE{Abeysekara2017science,
       author = {{Abeysekara}, A.~U. and {Albert}, A. and {Alfaro}, R. and {Alvarez}, C. and {{\'A}lvarez}, J.~D. and {Arceo}, R. and {Arteaga-Vel{\'a}zquez}, J.~C. and {Avila Rojas}, D. and {Ayala Solares}, H.~A. and {Barber}, A.~S. and {Bautista-Elivar}, N. and {Becerril}, A. and {Belmont-Moreno}, E. and {BenZvi}, S.~Y. and {Berley}, D. and {Bernal}, A. and {Braun}, J. and {Brisbois}, C. and {Caballero-Mora}, K.~S. and {Capistr{\'a}n}, T. and {Carrami{\~n}ana}, A. and {Casanova}, S. and {Castillo}, M. and {Cotti}, U. and {Cotzomi}, J. and {Couti{\~n}o de Le{\'o}n}, S. and {De Le{\'o}n}, C. and {De la Fuente}, E. and {Dingus}, B.~L. and {DuVernois}, M.~A. and {D{\'\i}az-V{\'e}lez}, J.~C. and {Ellsworth}, R.~W. and {Engel}, K. and {Enr{\'\i}quez-Rivera}, O. and {Fiorino}, D.~W. and {Fraija}, N. and {Garc{\'\i}a-Gonz{\'a}lez}, J.~A. and {Garfias}, F. and {Gerhardt}, M. and {Gonz{\'a}lez Mu{\~n}oz}, A. and {Gonz{\'a}lez}, M.~M. and {Goodman}, J.~A. and {Hampel-Arias}, Z. and {Harding}, J.~P. and {Hern{\'a}ndez}, S. and {Hern{\'a}ndez-Almada}, A. and {Hinton}, J. and {Hona}, B. and {Hui}, C.~M. and {H{\"u}ntemeyer}, P. and {Iriarte}, A. and {Jardin-Blicq}, A. and {Joshi}, V. and {Kaufmann}, S. and {Kieda}, D. and {Lara}, A. and {Lauer}, R.~J. and {Lee}, W.~H. and {Lennarz}, D. and {Vargas}, H. Le{\'o}n and {Linnemann}, J.~T. and {Longinotti}, A.~L. and {Luis Raya}, G. and {Luna-Garc{\'\i}a}, R. and {L{\'o}pez-Coto}, R. and {Malone}, K. and {Marinelli}, S.~S. and {Martinez}, O. and {Martinez-Castellanos}, I. and {Mart{\'\i}nez-Castro}, J. and {Mart{\'\i}nez-Huerta}, H. and {Matthews}, J.~A. and {Miranda-Romagnoli}, P. and {Moreno}, E. and {Mostaf{\'a}}, M. and {Nellen}, L. and {Newbold}, M. and {Nisa}, M.~U. and {Noriega-Papaqui}, R. and {Pelayo}, R. and {Pretz}, J. and {P{\'e}rez-P{\'e}rez}, E.~G. and {Ren}, Z. and {Rho}, C.~D. and {Rivi{\`e}re}, C. and {Rosa-Gonz{\'a}lez}, D. and {Rosenberg}, M. and {Ruiz-Velasco}, E. and {Salazar}, H. and {Salesa Greus}, F. and {Sandoval}, A. and {Schneider}, M. and {Schoorlemmer}, H. and {Sinnis}, G. and {Smith}, A.~J. and {Springer}, R.~W. and {Surajbali}, P. and {Taboada}, I. and {Tibolla}, O. and {Tollefson}, K. and {Torres}, I. and {Ukwatta}, T.~N. and {Vianello}, G. and {Weisgarber}, T. and {Westerhoff}, S. and {Wisher}, I.~G. and {Wood}, J. and {Yapici}, T. and {Yodh}, G. and {Younk}, P.~W. and {Zepeda}, A. and {Zhou}, H. and {Guo}, F. and {Hahn}, J. and {Li}, H. and {Zhang}, H.},
        title = "{Extended gamma-ray sources around pulsars constrain the origin of the positron flux at Earth}",
      journal = {Science},
         year = 2017,
        month = nov,
       volume = {358},
       number = {6365},
        pages = {911-914},
          doi = {10.1126/science.aan4880},
archivePrefix = {arXiv},
       eprint = {1711.06223},
 primaryClass = {astro-ph.HE},
       adsurl = {https://ui.adsabs.harvard.edu/abs/2017Sci...358..911A}
}

@ARTICLE{Sudoh2019,
       author = {{Sudoh}, Takahiro and {Linden}, Tim and {Beacom}, John F.},
        title = "{TeV halos are everywhere: Prospects for new discoveries}",
      journal = {\prd},
         year = 2019,
        month = aug,
       volume = {100},
       number = {4},
          eid = {043016},
        pages = {043016},
          doi = {10.1103/PhysRevD.100.043016},
archivePrefix = {arXiv},
       eprint = {1902.08203},
 primaryClass = {astro-ph.HE},
       adsurl = {https://ui.adsabs.harvard.edu/abs/2019PhRvD.100d3016S}
}

@ARTICLE{Evoli2018,
       author = {{Evoli}, Carmelo and {Linden}, Tim and {Morlino}, Giovanni},
        title = "{Self-generated cosmic-ray confinement in TeV halos: Implications for TeV {\ensuremath{\gamma}} -ray emission and the positron excess}",
      journal = {\prd},
         year = 2018,
        month = sep,
       volume = {98},
       number = {6},
          eid = {063017},
        pages = {063017},
          doi = {10.1103/PhysRevD.98.063017},
archivePrefix = {arXiv},
       eprint = {1807.09263},
 primaryClass = {astro-ph.HE},
       adsurl = {https://ui.adsabs.harvard.edu/abs/2018PhRvD..98f3017E}
}

@ARTICLE{Albert2024,
       author = {{Albert}, A. and {Alfaro}, R. and {Alvarez}, C. and {Arteaga-Vel{\'a}zquez}, J.~C. and {Avila Rojas}, D. and {Ayala Solares}, H.~A. and {Babu}, R. and {Belmont-Moreno}, E. and {Bernal}, A. and {Caballero-Mora}, K.~S. and {Capistr{\'a}n}, T. and {Carrami{\~n}ana}, A. and {Casanova}, S. and {Cotti}, U. and {Cotzomi}, J. and {Couti{\~n}o de Le{\'o}n}, S. and {de la Fuente}, E. and {Depaoli}, D. and {Di Lalla}, N. and {Diaz Hernandez}, R. and {Dingus}, B.~L. and {DuVernois}, M.~A. and {Durocher}, M. and {D{\'\i}az-V{\'e}lez}, J.~C. and {Engel}, K. and {Espinoza}, C. and {Fan}, K.~L. and {Fang}, K. and {Fraija}, N. and {Garc{\'\i}a-Gonz{\'a}lez}, J.~A. and {Garfias}, F. and {Goksu}, H. and {Gonz{\'a}lez}, M.~M. and {Goodman}, J.~A. and {Groetsch}, S. and {Harding}, J.~P. and {Hern{\'a}ndez-Cadena}, S. and {Herzog}, I. and {H{\"u}ntemeyer}, P. and {Huang}, D. and {Hueyotl-Zahuantitla}, F. and {Iriarte}, A. and {Joshi}, V. and {Kaufmann}, S. and {Kieda}, D. and {Lara}, A. and {Lee}, W.~H. and {Lee}, J. and {Le{\'o}n Vargas}, H. and {Linnemann}, J.~T. and {Longinotti}, A.~L. and {Luis-Raya}, G. and {Malone}, K. and {Martinez}, O. and {Mart{\'\i}nez-Castro}, J. and {Matthews}, J.~A. and {Miranda-Romagnoli}, P. and {Montes}, J.~A. and {Morales-Soto}, J.~A. and {Moreno}, E. and {Mostaf{\'a}}, M. and {Nayerhoda}, A. and {Nellen}, L. and {Noriega-Papaqui}, R. and {Olivera-Nieto}, L. and {Omodei}, N. and {P{\'e}rez Araujo}, Y. and {P{\'e}rez-P{\'e}rez}, E.~G. and {Rho}, C.~D. and {Rosa-Gonz{\'a}lez}, D. and {Salazar}, H. and {Salazar-Gallegos}, D. and {Sandoval}, A. and {Schneider}, M. and {Schwefer}, G. and {Serna-Franco}, J. and {Son}, Y. and {Springer}, R.~W. and {Tibolla}, O. and {Tollefson}, K. and {Torres}, I. and {Torres-Escobedo}, R. and {Turner}, R. and {Urea-Mena}, F. and {Varela}, E. and {Villase{\~n}or}, L. and {Wang}, X. and {Watson}, I.~J. and {Willox}, E. and {Wu}, H. and {Yun-C{\'a}rcamo}, S. and {Zhou}, H. and {de Le{\'o}n}, C. and {Di Mauro}, M.},
        title = "{Precise Measurements of TeV Halos around Geminga and Monogem Pulsars with HAWC}",
      journal = {\apj},
         year = 2024,
        month = oct,
       volume = {974},
       number = {2},
          eid = {246},
        pages = {246},
          doi = {10.3847/1538-4357/ad738e},
       adsurl = {https://ui.adsabs.harvard.edu/abs/2024ApJ...974..246A}
}

@ARTICLE{Adams2025,
       author = {{Adams}, C.~B. and {Archer}, A. and {Bangale}, P. and {Bartkoske}, J.~T. and {Benbow}, W. and {Buckley}, J.~H. and {Chen}, Y. and {Christiansen}, J.~L. and {Chromey}, A.~J. and {Duerr}, A. and {Errando}, M. and {Escobar Godoy}, M. and {Falcone}, A. and {Feldman}, S. and {Feng}, Q. and {Fortson}, L. and {Furniss}, A. and {Hanlon}, W. and {Hervet}, O. and {Hinrichs}, C.~E. and {Holder}, J. and {Humensky}, T.~B. and {Jin}, W. and {Johnson}, M.~N. and {Kaaret}, P. and {Kertzman}, M. and {Kherlakian}, M. and {Kieda}, D. and {Kleiner}, T.~K. and {Korzoun}, N. and {Krennrich}, F. and {Kumar}, S. and {Kundu}, S. and {Lang}, M.~J. and {Lundy}, M. and {Maier}, G. and {Millard}, M.~J. and {Millis}, J. and {Mooney}, C.~L. and {Moriarty}, P. and {Mukherjee}, R. and {Ning}, W. and {Ong}, R.~A. and {Pandey}, A. and {Pohl}, M. and {Pueschel}, E. and {Quinn}, J. and {Rabinowitz}, P.~L. and {Ragan}, K. and {Reynolds}, P.~T. and {Ribeiro}, D. and {Rizk}, L. and {Roache}, E. and {Sadeh}, I. and {Saha}, L. and {Sembroski}, G.~H. and {Shang}, R. and {Splettstoesser}, M. and {Tak}, D. and {Talluri}, A.~K. and {Tucci}, J.~V. and {Valverde}, J. and {Williams}, D.~A. and {Wong}, S.~L. and {Woo}, J. and {VERITAS Collaboration} and {Kwong}, Jon and {Mori}, Kaya and {Hailey}, Charles J. and {Safi-Harb}, Samar and {Zhang}, Shuo and {Tsuji}, Naomi and {XMM-Newton Collaboration} and {Manconi}, Silvia and {Donato}, Fiorenza and {di Mauro}, Mattia},
        title = "{Multiwavelength Observation of a Candidate Pulsar Halo LHAASO J0621+3755 and the First X-Ray Detection of PSR J0622+3749}",
      journal = {\apj},
         year = 2025,
        month = may,
       volume = {985},
       number = {1},
          eid = {90},
        pages = {90},
          doi = {10.3847/1538-4357/adc92b},
archivePrefix = {arXiv},
       eprint = {2504.02185},
 primaryClass = {astro-ph.HE},
       adsurl = {https://ui.adsabs.harvard.edu/abs/2025ApJ...985...90A}
}

@ARTICLE{HESS2005Science,
       author = {{Aharonian}, F. and {Akhperjanian}, A.~G. and {Aye}, K.-M. and {Bazer-Bachi}, A.~R. and {Beilicke}, M. and {Benbow}, W. and {Berge}, D. and {Berghaus}, P. and {Bernl{\"o}hr}, K. and {Boisson}, C. and {Bolz}, O. and {Borgmeier}, C. and {Braun}, I. and {Breitling}, F. and {Brown}, A.~M. and {Gordo}, J. Bussons and {Chadwick}, P.~M. and {Chounet}, L.-M. and {Cornils}, R. and {Costamante}, L. and {Degrange}, B. and {Djannati-Ata{\"\i}}, A. and {Drury}, L. O'C. and {Dubus}, G. and {Ergin}, T. and {Espigat}, P. and {Feinstein}, F. and {Fleury}, P. and {Fontaine}, G. and {Funk}, S. and {Gallant}, Y.~A. and {Giebels}, B. and {Gillessen}, S. and {Goret}, P. and {Hadjichristidis}, C. and {Hauser}, M. and {Heinzelmann}, G. and {Henri}, G. and {Hermann}, G. and {Hinton}, J.~A. and {Hofmann}, W. and {Holleran}, M. and {Horns}, D. and {de Jager}, O.~C. and {Jung}, I. and {Kh{\'e}lifi}, B. and {Komin}, Nu. and {Konopelko}, A. and {Latham}, I.~J. and {Le Gallou}, R. and {Lemi{\`e}re}, A. and {Lemoine}, M. and {Leroy}, N. and {Lohse}, T. and {Marcowith}, A. and {Masterson}, C. and {McComb}, T.~J.~L. and {de Naurois}, M. and {Nolan}, S.~J. and {Noutsos}, A. and {Orford}, K.~J. and {Osborne}, J.~L. and {Ouchrif}, M. and {Panter}, M. and {Pelletier}, G. and {Pita}, S. and {P{\"u}hlhofer}, G. and {Punch}, M. and {Raubenheimer}, B.~C. and {Raue}, M. and {Raux}, J. and {Rayner}, S.~M. and {Redondo}, I. and {Reimer}, A. and {Reimer}, O. and {Ripken}, J. and {Rob}, L. and {Rolland}, L. and {Rowell}, G. and {Sahakian}, V. and {Saug{\'e}}, L. and {Schlenker}, S. and {Schlickeiser}, R. and {Schuster}, C. and {Schwanke}, U. and {Siewert}, M. and {Sol}, H. and {Steenkamp}, R. and {Stegmann}, C. and {Tavernet}, J.-P. and {Terrier}, R. and {Th{\'e}oret}, C.~G. and {Tluczykont}, M. and {van der Walt}, D.~J. and {Vasileiadis}, G. and {Venter}, C. and {Vincent}, P. and {Visser}, B. and {V{\"o}lk}, H.~J. and {Wagner}, S.~J.},
        title = "{A New Population of Very High Energy Gamma-Ray Sources in the Milky Way}",
      journal = {Science},
         year = 2005,
        month = mar,
       volume = {307},
       number = {5717},
        pages = {1938-1942},
          doi = {10.1126/science.1108643},
archivePrefix = {arXiv},
       eprint = {astro-ph/0504380},
 primaryClass = {astro-ph},
       adsurl = {https://ui.adsabs.harvard.edu/abs/2005Sci...307.1938A}
}

@ARTICLE{HESS2024,
       author = {{H.~E.~S.~S. Collaboration} and {Aharonian}, F. and {Ait Benkhali}, F. and {Aschersleben}, J. and {Ashkar}, H. and {Backes}, M. and {Baktash}, A. and {Barbosa Martins}, V. and {Barnard}, J. and {Batzofin}, R. and {Becherini}, Y. and {Berge}, D. and {Bernl{\"o}hr}, K. and {Bi}, B. and {B{\"o}ttcher}, M. and {Boisson}, C. and {Bolmont}, J. and {de Bony de Lavergne}, M. and {Borowska}, J. and {Bouyahiaoui}, M. and {Breuhaus}, M. and {Brose}, R. and {Brun}, F. and {Bruno}, B. and {Bulik}, T. and {Burger-Scheidlin}, C. and {Caroff}, S. and {Casanova}, S. and {Cecil}, R. and {Celic}, J. and {Cerruti}, M. and {Chambery}, P. and {Chand}, T. and {Chen}, A. and {Chibueze}, J. and {Chibueze}, O. and {Cotter}, G. and {Damascene Mbarubucyeye}, J. and {Djannati-Ata{\"\i}}, A. and {Dmytriiev}, A. and {Doroshenko}, V. and {Einecke}, S. and {Ernenwein}, J.-P. and {Feijen}, K. and {Filipovic}, M. and {Fontaine}, G. and {F{\"u}{\ss}ling}, M. and {Funk}, S. and {Gabici}, S. and {Gallant}, Y.~A. and {Ghafourizadeh}, S. and {Giavitto}, G. and {Glawion}, D. and {Glicenstein}, J.~F. and {Goswami}, P. and {Grolleron}, G. and {Grondin}, M.-H. and {Hinton}, J.~A. and {Hofmann}, W. and {Holch}, T.~L. and {Holler}, M. and {Jamrozy}, M. and {Jankowsky}, F. and {Joshi}, V. and {Jung-Richardt}, I. and {Katarzy{\'n}ski}, K. and {Khatoon}, R. and {Kh{\'e}lifi}, B. and {Klepser}, S. and {Klu{\'z}niak}, W. and {Komin}, Nu. and {Kosack}, K. and {Kostunin}, D. and {Kundu}, A. and {Lang}, R.~G. and {Le Stum}, S. and {Leitl}, F. and {Lemi{\`e}re}, A. and {Lemoine-Goumard}, M. and {Lenain}, J.-P. and {Leuschner}, F. and {Mackey}, J. and {Malyshev}, D. and {Malyshev}, D. and {Marandon}, V. and {Marinos}, P. and {Mart{\'\i}-Devesa}, G. and {Marx}, R. and {Mehta}, A. and {Mitchell}, A. and {Moderski}, R. and {Mohrmann}, L. and {Montanari}, A. and {Moulin}, E. and {Murach}, T. and {de Naurois}, M. and {Niemiec}, J. and {Priyana Noel}, A. and {O'Brien}, P. and {Ohm}, S. and {Olivera-Nieto}, L. and {de Ona Wilhelmi}, E. and {Ostrowski}, M. and {Panny}, S. and {Panter}, M. and {Parsons}, R.~D. and {Prokhorov}, D.~A. and {P{\"u}hlhofer}, G. and {Punch}, M. and {Quirrenbach}, A. and {Regeard}, M. and {Reichherzer}, P. and {Reimer}, A. and {Reimer}, O. and {Ren}, H. and {Renaud}, M. and {Reville}, B. and {Rieger}, F. and {Roellinghoff}, G. and {Rudak}, B. and {Sahakian}, V. and {Salzmann}, H. and {Sasaki}, M. and {Sch{\"u}ssler}, F. and {Schutte}, H.~M. and {Shapopi}, J.~N.~S. and {Specovius}, A. and {Spencer}, S. and {Steenkamp}, R. and {Steinmassl}, S. and {Steppa}, C. and {Sushch}, I. and {Suzuki}, H. and {Takahashi}, T. and {Tanaka}, T. and {Terrier}, R. and {Tluczykont}, M. and {Tsuji}, N. and {Unbehaun}, T. and {van Eldik}, C. and {Vecchi}, M. and {Veh}, J. and {Venter}, C. and {Vink}, J. and {Wach}, T. and {Wagner}, S.~J. and {Wierzcholska}, A. and {Zacharias}, M. and {Zargaryan}, D. and {Zdziarski}, A.~A. and {Zouari}, S. and {{\.Z}ywucka}, N.},
        title = "{Unveiling extended gamma-ray emission around HESS J1813-178}",
      journal = {\aap},
         year = 2024,
        month = jun,
       volume = {686},
          eid = {A149},
        pages = {A149},
          doi = {10.1051/0004-6361/202348374},
archivePrefix = {arXiv},
       eprint = {2403.16802},
 primaryClass = {astro-ph.HE},
       adsurl = {https://ui.adsabs.harvard.edu/abs/2024A&A...686A.149H}
}

@ARTICLE{Fang2010,
       author = {{Fang}, Jun and {Zhang}, Li},
        title = "{Multiband Nonthermal Radiative Properties of HESS J1813-178}",
      journal = {\apj},
         year = 2010,
        month = jul,
       volume = {718},
       number = {1},
        pages = {467-473},
          doi = {10.1088/0004-637X/718/1/467},
archivePrefix = {arXiv},
       eprint = {1005.3386},
 primaryClass = {astro-ph.HE},
       adsurl = {https://ui.adsabs.harvard.edu/abs/2010ApJ...718..467F}
}

@ARTICLE{Araya2018,
       author = {{Araya}, Miguel},
        title = "{GeV Emission in the Region of HESS J1809-193 and HESS J1813-178: Is HESS J1809-193 a Proton Pevatron?}",
      journal = {\apj},
         year = 2018,
        month = may,
       volume = {859},
       number = {1},
          eid = {69},
        pages = {69},
          doi = {10.3847/1538-4357/aabd7e},
archivePrefix = {arXiv},
       eprint = {1804.03325},
 primaryClass = {astro-ph.HE},
       adsurl = {https://ui.adsabs.harvard.edu/abs/2018ApJ...859...69A}
}

@ARTICLE{Guo2024,
       author = {{Guo}, Xiaolei and {Xin}, Yuliang},
        title = "{Revision of the GeV {\ensuremath{\gamma}}-Ray Emission in the Region of HESS J1813-178 with Fermi-LAT}",
      journal = {\apj},
         year = 2024,
        month = apr,
       volume = {965},
       number = {1},
          eid = {28},
        pages = {28},
          doi = {10.3847/1538-4357/ad2ae1},
archivePrefix = {arXiv},
       eprint = {2402.11880},
 primaryClass = {astro-ph.HE},
       adsurl = {https://ui.adsabs.harvard.edu/abs/2024ApJ...965...28G}
}

@ARTICLE{Schwefer2025,
       author = {{Schwefer}, Georg and {Hinton}, Jim},
        title = "{High-Resolution Measurements with the CTAO Southern Array: The Case for Pulsar Wind Nebulae}",
      journal = {arXiv e-prints},
         year = 2025,
        month = dec,
          eid = {arXiv:2512.17498},
        pages = {arXiv:2512.17498},
          doi = {10.48550/arXiv.2512.17498},
archivePrefix = {arXiv},
       eprint = {2512.17498},
 primaryClass = {astro-ph.HE},
       adsurl = {https://ui.adsabs.harvard.edu/abs/2025arXiv251217498S}
}

@ARTICLE{Xian2026,
       author = {{Xian}, Yujun and {Yu}, Huan and {Zhou}, Liancheng and {Fang}, Jun},
        title = "{The PWN─TeV Halo Transition in HESS J1809─193 Revealed by a Multizone Transport Model}",
      journal = {\apj},
         year = 2026,
        month = feb,
       volume = {997},
       number = {2},
          eid = {213},
        pages = {213},
          doi = {10.3847/1538-4357/ae2851},
       adsurl = {https://ui.adsabs.harvard.edu/abs/2026ApJ...997..213X}
}

@ARTICLE{Brogan2005,
       author = {{Brogan}, C.~L. and {Gaensler}, B.~M. and {Gelfand}, J.~D. and {Lazendic}, J.~S. and {Lazio}, T.~J.~W. and {Kassim}, N.~E. and {McClure-Griffiths}, N.~M.},
        title = "{Discovery of a Radio Supernova Remnant and Nonthermal X-Rays Coincident with the TeV Source HESS J1813-178}",
      journal = {\apjl},
         year = 2005,
        month = aug,
       volume = {629},
       number = {2},
        pages = {L105-L108},
          doi = {10.1086/491471},
archivePrefix = {arXiv},
       eprint = {astro-ph/0505145},
 primaryClass = {astro-ph},
       adsurl = {https://ui.adsabs.harvard.edu/abs/2005ApJ...629L.105B}
}

@ARTICLE{Wach2023,
       author = {{Wach}, T. and {Mitchell}, A.~M.~W. and {Joshi}, V. and {Funk}, S.},
        title = "{Joint H.E.S.S. and Fermi-LAT analysis of the region around PSR J1813-1749}",
      journal = {arXiv e-prints},
         year = 2023,
        month = aug,
          eid = {arXiv:2308.16717},
        pages = {arXiv:2308.16717},
          doi = {10.48550/arXiv.2308.16717},
archivePrefix = {arXiv},
       eprint = {2308.16717},
 primaryClass = {astro-ph.HE},
       adsurl = {https://ui.adsabs.harvard.edu/abs/2023arXiv230816717W}
}

@ARTICLE{van2018arXiv,
       author = {{van Rensburg}, Carlo and {Venter}, Christo and {Kruger}, Paulus},
        title = "{Spectral Modelling of H.E.S.S.-detected Pulsar Wind Nebulae}",
      journal = {arXiv e-prints},
         year = 2018,
        month = feb,
          eid = {arXiv:1802.00216},
        pages = {arXiv:1802.00216},
          doi = {10.48550/arXiv.1802.00216},
archivePrefix = {arXiv},
       eprint = {1802.00216},
 primaryClass = {astro-ph.HE},
       adsurl = {https://ui.adsabs.harvard.edu/abs/2018arXiv180200216V}
}

@ARTICLE{Lu2017ApJ,
       author = {{Lu}, Fang-Wu and {Gao}, Quan-Gui and {Zhang}, Li},
        title = "{A Self-consistent and Spatially Dependent Model of the Multiband Emission of Pulsar Wind Nebulae}",
      journal = {\apj},
         year = 2017,
        month = jan,
       volume = {834},
       number = {1},
          eid = {43},
        pages = {43},
          doi = {10.3847/1538-4357/834/1/43},
       adsurl = {https://ui.adsabs.harvard.edu/abs/2017ApJ...834...43L}
}

@ARTICLE{Sergio2021,
       author = {{Dzib}, Sergio A. and {Rodr{\'\i}guez}, Luis F.},
        title = "{Radio Proper Motions of the Energetic Pulsar PSR J1813-1749}",
      journal = {\apj},
         year = 2021,
        month = dec,
       volume = {923},
       number = {2},
          eid = {228},
        pages = {228},
          doi = {10.3847/1538-4357/ac312f},
archivePrefix = {arXiv},
       eprint = {2110.09727},
 primaryClass = {astro-ph.HE},
       adsurl = {https://ui.adsabs.harvard.edu/abs/2021ApJ...923..228D}
}

@ARTICLE{Vorster2013,
       author = {{Vorster}, M.~J. and {Moraal}, H.},
        title = "{The Effect of Diffusion on the Particle Spectra in Pulsar Wind Nebulae}",
      journal = {\apj},
         year = 2013,
        month = mar,
       volume = {765},
       number = {1},
          eid = {30},
        pages = {30},
          doi = {10.1088/0004-637X/765/1/30},
archivePrefix = {arXiv},
       eprint = {1309.5813},
 primaryClass = {astro-ph.HE},
       adsurl = {https://ui.adsabs.harvard.edu/abs/2013ApJ...765...30V}
}

@ARTICLE{Ishizaki2018,
       author = {{Ishizaki}, Wataru and {Asano}, Katsuaki and {Kawaguchi}, Kyohei},
        title = "{Outflow and Emission Model of Pulsar Wind Nebulae with the Back Reaction of Particle Diffusion}",
      journal = {\apj},
         year = 2018,
        month = nov,
       volume = {867},
       number = {2},
          eid = {141},
        pages = {141},
          doi = {10.3847/1538-4357/aae389},
archivePrefix = {arXiv},
       eprint = {1809.09054},
 primaryClass = {astro-ph.HE},
       adsurl = {https://ui.adsabs.harvard.edu/abs/2018ApJ...867..141I}
}

@ARTICLE{Kundu2024,
       author = {{Kundu}, A. and {Joshi}, Jagdish C. and {Venter}, C. and {Engelbrecht}, N.~E. and {Zhang}, W. and {Torres}, Diego F. and {Sushch}, I. and {Tanaka}, Shuta J.},
        title = "{Spatio-spectral-temporal modelling of two young pulsar wind nebulae}",
      journal = {\mnras},
         year = 2024,
        month = dec,
       volume = {535},
       number = {3},
        pages = {2415-2435},
          doi = {10.1093/mnras/stae2435},
archivePrefix = {arXiv},
       eprint = {2410.18386},
 primaryClass = {astro-ph.HE},
       adsurl = {https://ui.adsabs.harvard.edu/abs/2024MNRAS.535.2415K}
}

@ARTICLE{Liu2019,
       author = {{Liu}, Ruo-Yu and {Ge}, Chong and {Sun}, Xiao-Na and {Wang}, Xiang-Yu},
        title = "{Constraining the Magnetic Field in the TeV Halo of Geminga with X-Ray Observations}",
      journal = {\apj},
         year = 2019,
        month = apr,
       volume = {875},
       number = {2},
          eid = {149},
        pages = {149},
          doi = {10.3847/1538-4357/ab125c},
archivePrefix = {arXiv},
       eprint = {1904.11438},
 primaryClass = {astro-ph.HE},
       adsurl = {https://ui.adsabs.harvard.edu/abs/2019ApJ...875..149L}
}

@ARTICLE{Zhu2018,
       author = {{Zhu}, Bo-Tao and {Zhang}, Li and {Fang}, Jun},
        title = "{Multiband nonthermal radiative properties of pulsar wind nebulae}",
      journal = {\aap},
         year = 2018,
        month = jan,
       volume = {609},
          eid = {A110},
        pages = {A110},
          doi = {10.1051/0004-6361/201629108},
       adsurl = {https://ui.adsabs.harvard.edu/abs/2018A&A...609A.110Z}
}

@ARTICLE{Holler2012,
       author = {{Holler}, M. and {Sch{\"o}ck}, F.~M. and {Eger}, P. and {Kie{\ss}ling}, D. and {Valerius}, K. and {Stegmann}, C.},
        title = "{Spatially resolved X-ray spectroscopy and modeling of the nonthermal emission of the pulsar wind nebula in G0.9+0.1}",
      journal = {\aap},
         year = 2012,
        month = mar,
       volume = {539},
          eid = {A24},
        pages = {A24},
          doi = {10.1051/0004-6361/201118121},
archivePrefix = {arXiv},
       eprint = {1201.2618},
 primaryClass = {astro-ph.GA},
       adsurl = {https://ui.adsabs.harvard.edu/abs/2012A&A...539A..24H}
}

@ARTICLE{Hnatyk2022,
       author = {{Hnatyk}, B. and {Hnatyk}, R. and {Zhdanov}, V. and {Voitsekhovskyi}, V.},
        title = "{Unveiling the nature of the unidentified gamma-ray sources 4FGL J1908.6+0915e, HESS J1907+089/HOTS J1907+091, and 3HWC J1907+085 in the sky region of the magnetar SGR 1900+14}",
      journal = {\mnras},
         year = 2022,
        month = jul,
       volume = {514},
       number = {1},
        pages = {762-779},
          doi = {10.1093/mnras/stac1304},
archivePrefix = {arXiv},
       eprint = {2009.06081},
 primaryClass = {astro-ph.HE},
       adsurl = {https://ui.adsabs.harvard.edu/abs/2022MNRAS.514..762H}
}

@ARTICLE{Zhang2008,
       author = {{Zhang}, L. and {Chen}, S.~B. and {Fang}, J.},
        title = "{Nonthermal Radiation from Pulsar Wind Nebulae}",
      journal = {\apj},
         year = 2008,
        month = apr,
       volume = {676},
       number = {2},
        pages = {1210-1217},
          doi = {10.1086/527466},
       adsurl = {https://ui.adsabs.harvard.edu/abs/2008ApJ...676.1210Z}
}

@ARTICLE{van2018MN,
       author = {{van Rensburg}, C. and {Kr{\"u}ger}, P.~P. and {Venter}, C.},
        title = "{Spatially dependent modelling of pulsar wind nebula G0.9+0.1}",
      journal = {\mnras},
         year = 2018,
        month = jul,
       volume = {477},
       number = {3},
        pages = {3853-3868},
          doi = {10.1093/mnras/sty826},
archivePrefix = {arXiv},
       eprint = {1803.10625},
 primaryClass = {astro-ph.HE},
       adsurl = {https://ui.adsabs.harvard.edu/abs/2018MNRAS.477.3853V}
}

@ARTICLE{Lu2017MN,
       author = {{Lu}, Fang-Wu and {Gao}, Quan-Gui and {Zhu}, Bo-Tao and {Zhang}, Li},
        title = "{Spatially radiative properties of 3C 58 and G21.5-0.9}",
      journal = {\mnras},
         year = 2017,
        month = dec,
       volume = {472},
       number = {3},
        pages = {2926-2936},
          doi = {10.1093/mnras/stx2223},
       adsurl = {https://ui.adsabs.harvard.edu/abs/2017MNRAS.472.2926L}
}

@ARTICLE{CFL1928,
       author = {{Courant}, R. and {Friedrichs}, K. and {Lewy}, H.},
        title = "{{\"U}ber die partiellen Differenzengleichungen der mathematischen Physik}",
      journal = {Mathematische Annalen},
         year = 1928,
        month = jan,
       volume = {100},
        pages = {32-74},
          doi = {10.1007/BF01448839},
       adsurl = {https://ui.adsabs.harvard.edu/abs/1928MatAn.100...32C}
}

@ARTICLE{Wu2023,
       author = {{Wu}, Keyao and {Zhou}, Liancheng and {Gong}, Yunlu and {Fang}, Jun},
        title = "{Investigating the radiative properties of LHAASO J1908 + 0621}",
      journal = {\mnras},
         year = 2023,
        month = feb,
       volume = {519},
       number = {2},
        pages = {1881-1886},
          doi = {10.1093/mnras/stac3618},
archivePrefix = {arXiv},
       eprint = {2212.04040},
 primaryClass = {astro-ph.HE},
       adsurl = {https://ui.adsabs.harvard.edu/abs/2023MNRAS.519.1881W}
}

@ARTICLE{Tian2023,
       author = {{Tian}, Shiting and {Zhou}, Liancheng and {Gong}, Yunlu and {Wu}, Keyao and {Fang}, Jun and {Yu}, Huan},
        title = "{Investigating the Multiband Nonthermal Radiative Properties of HESS J1420-607}",
      journal = {\pasp},
         year = 2023,
        month = jul,
       volume = {135},
       number = {1049},
          eid = {074503},
        pages = {074503},
          doi = {10.1088/1538-3873/ace3f8},
       adsurl = {https://ui.adsabs.harvard.edu/abs/2023PASP..135g4503T}
}

@ARTICLE{Halpern2012,
       author = {{Halpern}, J.~P. and {Gotthelf}, E.~V. and {Camilo}, F.},
        title = "{Spin-down Measurement of PSR J1813-1749: The Energetic Pulsar Powering HESS J1813-178}",
      journal = {\apjl},
         year = 2012,
        month = jul,
       volume = {753},
       number = {1},
          eid = {L14},
        pages = {L14},
          doi = {10.1088/2041-8205/753/1/L14},
archivePrefix = {arXiv},
       eprint = {1206.2338},
 primaryClass = {astro-ph.HE},
       adsurl = {https://ui.adsabs.harvard.edu/abs/2012ApJ...753L..14H}
}

@ARTICLE{Camilo2021,
       author = {{Camilo}, F. and {Ransom}, S.~M. and {Halpern}, J.~P. and {Roshi}, D. Anish},
        title = "{Radio Detection of PSR J1813-1749 in HESS J1813-178: The Most Scattered Pulsar Known}",
      journal = {\apj},
         year = 2021,
        month = aug,
       volume = {917},
       number = {2},
          eid = {67},
        pages = {67},
          doi = {10.3847/1538-4357/ac0720},
archivePrefix = {arXiv},
       eprint = {2106.00386},
 primaryClass = {astro-ph.HE},
       adsurl = {https://ui.adsabs.harvard.edu/abs/2021ApJ...917...67C}
}

@INPROCEEDINGS{Zabalza2015,
       author = {{Zabalza}, V.},
        title = "{Naima: a Python package for inference of particle distribution properties from nonthermal spectra}",
    booktitle = {34th International Cosmic Ray Conference (ICRC2015)},
         year = 2015,
       series = {International Cosmic Ray Conference},
       volume = {34},
        month = jul,
          eid = {922},
        pages = {922},
          doi = {10.22323/1.236.0922},
archivePrefix = {arXiv},
       eprint = {1509.03319},
 primaryClass = {astro-ph.HE},
       adsurl = {https://ui.adsabs.harvard.edu/abs/2015ICRC...34..922Z}
}

@ARTICLE{Torres2014,
       author = {{Torres}, D.~F. and {Cillis}, A. and {Mart{\'\i}n}, J. and {de O{\~n}a Wilhelmi}, E.},
        title = "{Time-dependent modeling of TeV-detected, young pulsar wind nebulae}",
      journal = {Journal of High Energy Astrophysics},
         year = 2014,
        month = may,
       volume = {1},
        pages = {31-62},
          doi = {10.1016/j.jheap.2014.02.001},
archivePrefix = {arXiv},
       eprint = {1402.5485},
 primaryClass = {astro-ph.HE},
       adsurl = {https://ui.adsabs.harvard.edu/abs/2014JHEAp...1...31T}
}

@ARTICLE{Wu2025,
       author = {{Wu}, Keyao and {Xiao}, Yifan and {Fang}, Jun and {Yu}, Huan},
        title = "{Multi-wavelength Non-thermal Radiative Properties of Pulsar Wind Nebulae with Ages Around 10 kyr}",
      journal = {\pasp},
         year = 2025,
        month = may,
       volume = {137},
       number = {5},
          eid = {054102},
        pages = {054102},
          doi = {10.1088/1538-3873/adcf58},
       adsurl = {https://ui.adsabs.harvard.edu/abs/2025PASP..137e4102W}
}

@ARTICLE{Ptuskin2006,
       author = {{Ptuskin}, Vladimir S. and {Moskalenko}, Igor V. and {Jones}, Frank C. and {Strong}, Andrew W. and {Zirakashvili}, Vladimir N.},
        title = "{Dissipation of Magnetohydrodynamic Waves on Energetic Particles: Impact on Interstellar Turbulence and Cosmic-Ray Transport}",
      journal = {\apj},
         year = 2006,
        month = may,
       volume = {642},
       number = {2},
        pages = {902-916},
          doi = {10.1086/501117},
archivePrefix = {arXiv},
       eprint = {astro-ph/0510335},
 primaryClass = {astro-ph},
       adsurl = {https://ui.adsabs.harvard.edu/abs/2006ApJ...642..902P}
}

@ARTICLE{Strong2007,
       author = {{Strong}, Andrew W. and {Moskalenko}, Igor V. and {Ptuskin}, Vladimir S.},
        title = "{Cosmic-Ray Propagation and Interactions in the Galaxy}",
      journal = {Annual Review of Nuclear and Particle Science},
         year = 2007,
        month = nov,
       volume = {57},
       number = {1},
        pages = {285-327},
          doi = {10.1146/annurev.nucl.57.090506.123011},
archivePrefix = {arXiv},
       eprint = {astro-ph/0701517},
 primaryClass = {astro-ph},
       adsurl = {https://ui.adsabs.harvard.edu/abs/2007ARNPS..57..285S}
}

@ARTICLE{Helfand2007,
       author = {{Helfand}, D.~J. and {Gotthelf}, E.~V. and {Halpern}, J.~P. and {Camilo}, F. and {Semler}, D.~R. and {Becker}, R.~H. and {White}, R.~L.},
        title = "{Discovery of the Putative Pulsar and Wind Nebula Associated with the TeV Gamma-Ray Source HESS J1813-178}",
      journal = {\apj},
         year = 2007,
        month = aug,
       volume = {665},
       number = {2},
        pages = {1297-1303},
          doi = {10.1086/519734},
archivePrefix = {arXiv},
       eprint = {0705.0065},
 primaryClass = {astro-ph},
       adsurl = {https://ui.adsabs.harvard.edu/abs/2007ApJ...665.1297H}
}

@ARTICLE{Funk2007,
       author = {{Funk}, S. and {Hinton}, J.~A. and {Moriguchi}, Y. and {Aharonian}, F.~A. and {Fukui}, Y. and {Hofmann}, W. and {Horns}, D. and {P{\"u}hlhofer}, G. and {Reimer}, O. and {Rowell}, G. and {Terrier}, R. and {Vink}, J. and {Wagner}, S.~J.},
        title = "{XMM-Newton observations of HESS J1813-178 reveal a composite Supernova remnant}",
      journal = {\aap},
         year = 2007,
        month = jul,
       volume = {470},
       number = {1},
        pages = {249-257},
          doi = {10.1051/0004-6361:20066779},
archivePrefix = {arXiv},
       eprint = {astro-ph/0611646},
 primaryClass = {astro-ph},
       adsurl = {https://ui.adsabs.harvard.edu/abs/2007A&A...470..249F}
}

@ARTICLE{Ubertini2005,
       author = {{Ubertini}, P. and {Bassani}, L. and {Malizia}, A. and {Bazzano}, A. and {Bird}, A.~J. and {Dean}, A.~J. and {De Rosa}, A. and {Lebrun}, F. and {Moran}, L. and {Renaud}, M. and {Stephen}, J.~B. and {Terrier}, R. and {Walter}, R.},
        title = "{INTEGRAL IGR J18135-1751 = HESS J1813-178: A New Cosmic High-Energy Accelerator from keV to TeV Energies}",
      journal = {\apjl},
         year = 2005,
        month = aug,
       volume = {629},
       number = {2},
        pages = {L109-L112},
          doi = {10.1086/447766},
archivePrefix = {arXiv},
       eprint = {astro-ph/0505191},
 primaryClass = {astro-ph},
       adsurl = {https://ui.adsabs.harvard.edu/abs/2005ApJ...629L.109U}
}

@ARTICLE{Gotthelf2009,
       author = {{Gotthelf}, E.~V. and {Halpern}, J.~P.},
        title = "{Discovery of a Highly Energetic X-Ray Pulsar Powering HESS J1813-178 in the Young Supernova Remnant G12.82-0.02}",
      journal = {\apjl},
         year = 2009,
        month = aug,
       volume = {700},
       number = {2},
        pages = {L158-L161},
          doi = {10.1088/0004-637X/700/2/L158},
archivePrefix = {arXiv},
       eprint = {0907.0137},
 primaryClass = {astro-ph.HE},
       adsurl = {https://ui.adsabs.harvard.edu/abs/2009ApJ...700L.158G}
}

@ARTICLE{Pacini1967,
       author = {{Pacini}, F.},
        title = "{Energy Emission from a Neutron Star}",
      journal = {\nat},
         year = 1967,
        month = nov,
       volume = {216},
       number = {5115},
        pages = {567-568},
          doi = {10.1038/216567a0},
       adsurl = {https://ui.adsabs.harvard.edu/abs/1967Natur.216..567P}
}

@book{Manchester1977,
  author    = {Manchester, R. N. and Taylor, J. H.},
  title     = {Pulsars},
  year      = {1977},
  publisher = {W. H. Freeman},
  address   = {San Francisco, CA}
}

@ARTICLE{Pacini1973,
       author = {{Pacini}, F. and {Salvati}, M.},
        title = "{On the Evolution of Supernova Remnants. Evolution of the Magnetic Field, Particles, Content, and Luminosity}",
      journal = {\apj},
         year = 1973,
        month = nov,
       volume = {186},
        pages = {249-266},
          doi = {10.1086/152495},
       adsurl = {https://ui.adsabs.harvard.edu/abs/1973ApJ...186..249P}
}

\end{document}